\documentclass[twocolumn]{aastex631}

\usepackage{pdfcomment}
\newif\ifhidedel\hidedelfalse
\newcommand*\editopc{}
\newcommand*\smallsig[2]{%
  \begingroup\raisebox{.7ex}{\color{#2}\sffamily\bfseries\scriptsize#1}\endgroup}
\newcommand*\largesig[2]{%
  \begingroup\color{#2}\sffamily\bfseries#1\endgroup~}
\newcommand\textdblbrk[1]{%
  \begingroup\normalfont[\kern-.15em[#1]\kern-.15em]\endgroup}
\newcommand*\DeclareAddCommand[2]{%
  \expandafter\newcommand\csname add#1\endcsname[1]{\texorpdfstring
    {\smallsig{#1}{#2!25}%
    \begingroup\color{#2\editopc}##1\endgroup}
    {##1}}}
\newcommand*\DeclareDeleteCommand[2]{%
  \expandafter\newcommand\csname del#1\endcsname[1]{\texorpdfstring
    {\ifhidedel\else
      \smallsig{#1}{#2!25}%
      \begingroup\def\editopc{!25}\color{#2\editopc}##1\endgroup
    \fi}
    {}}}
\newcommand*\DeclareReplaceCommand[2]{%
  \expandafter\newcommand\csname repl#1\endcsname[2]{\texorpdfstring
    {\smallsig{#1}{#2!25}%
    \ifhidedel\else
      \begingroup\def\editopc{!25}\color{#2\editopc}##1\endgroup
    \fi
    \begingroup\color{#2\editopc}##2\endgroup}
    {##2}}}
\newcommand*\DeclareNoteCommand[2]{%
  \expandafter\newcommand\csname note#1\endcsname[1]{\texorpdfstring
    {\ifhidedel\else
      \begingroup\color{#2}\textdblbrk{\largesig{#1}{#2}##1}\endgroup
    \fi}
    {}}}
\newcommand*\DeclarePopupNoteCommand[2]{%
  \expandafter\newcommand\csname popup#1\endcsname[1]{\texorpdfstring{%
    \begingroup\color{#2\editopc}%
    \normalfont\sffamily\footnotesize
    \pdfmarkupcomment[markup=Highlight, color=yellow!33!white, author=#1]
      {#1}{##1}%
    \endgroup
  }{}}}
\newcommand*\DeclareEditCommands[2]{%
  \DeclareAddCommand{#1}{#2}%
  \DeclareDeleteCommand{#1}{#2}%
  \DeclareReplaceCommand{#1}{#2}%
  \DeclareNoteCommand{#1}{#2}
  \DeclarePopupNoteCommand{#1}{#2}}

\DeclareEditCommands{EC}{green!70!black}
\DeclareEditCommands{TB}{magenta}

\usepackage{CJK}
\usepackage{amsmath}

\shorttitle{RMHD Simulation of sub-Eddington Circumbinary Disk in 10:1 MBHB}
\shortauthors{Tiwari et al.}

\usepackage{CJK}

\usepackage{gensymb}

\begin{document}
\defcitealias{tiwariRMHD}{Paper~I}
\begin{CJK*}{UTF8}{gbsn}
\title{Radiation Magnetohydrodynamic Simulation of sub-Eddington Circumbinary Disk around a q=0.1 Massive Black Hole Binary}

\title{Radiation Magnetohydrodynamic Simulation of sub-Eddington Circumbinary Disk in a 10:1 Massive Black Hole Binary}

 \correspondingauthor{Vishal Tiwari}
 \email{vtiwari@gatech.edu}

  \author[0000-0002-7110-9885]{Vishal Tiwari}
  \author[0000-0001-5949-6109]{Chi-Ho Chan}
  \author[0000-0002-7835-7814]{Tamara Bogdanovi\'c}
  \affiliation{Center for Relativistic Astrophysics and School of Physics,
  Georgia Institute of Technology, Atlanta, GA 30332, USA}
  \author[0000-0002-2624-3399]{Yan-Fei Jiang (姜燕飞)}
  \affiliation{Center for Computational Astrophysics, Flatiron Institute,
  162 Fifth Avenue, New York, NY 10010, USA}
  \author[0000-0001-7488-4468]{Shane W. Davis}
  \affiliation{Department of Astronomy, University of Virginia,
  Charlottesville, VA 22904, USA}

%% Note that the \and command from previous versions of AASTeX is now
%% depreciated in this version as it is no longer necessary. AASTeX 
%% automatically takes care of all commas and "and"s between authors names.

%% AASTeX 6.31 has the new \collaboration and \nocollaboration commands to
%% provide the collaboration status of a group of authors. These commands 
%% can be used either before or after the list of corresponding authors. The
%% argument for \collaboration is the collaboration identifier. Authors are
%% encouraged to surround collaboration identifiers with ()s. The 
%% \nocollaboration command takes no argument and exists to indicate that
%% the nearby authors are not part of surrounding collaborations.

%% Mark off the abstract in the ``abstract'' environment. 

\begin{abstract}
We present a global three-dimensional radiation magnetohydrodynamic (RMHD) simulation of a circumbinary disk (CBD) around a massive black hole binary (MBHB) with a total mass $2 \times 10^7\,M_{\odot}$ and mass ratio $0.1$, separated by $100\, GM_{\rm tot}/c^2$. The inclusion of radiation makes the disk thinner, denser, less eccentric at the inner edge, and more filamentary when compared to an otherwise identical locally isothermal MHD disk. The RMHD disk has accretion rate $\sim 0.23\,\dot{M}_{\mathrm{Edd}}$ and produces thermal emission peaking in the near-UV/optical with a luminosity of $\sim 1\, \% L_{\rm {Edd }}$. Compared with an equal-mass binary with the same total mass, the thermal emission of the CBD around the unequal-mass binary is several orders of magnitude brighter and much more variable at far-UV/soft X-rays frequencies. Similarly, we find that the light curve associated with the $0.1$ mass ratio binary exhibits dominant periodicity corresponding to 2 binary orbits, compared to the equal-mass binary that shows periodicity at 2.5-5 binary orbits. Our results highlight the importance of radiation for the structure and observational properties of MBHB circumbinary disks and have implications for detecting electromagnetic counterparts to LISA gravitational wave precursors and for the heavier binaries targeted by the Pulsar Timing Arrays.
\end{abstract}

%% Keywords should appear after the \end{abstract} command. 
%% The AAS Journals now uses Unified Astronomy Thesaurus concepts:
%% https://astrothesaurus.org
%% You will be asked to selected these concepts during the submission process
%% but this old "keyword" functionality is maintained in case authors want
%% to include these concepts in their preprints.
\keywords{Radiative magnetohydrodynamics (2009) --- Supermassive black holes (1663) --- Gravitational wave sources (677) ---
  Accretion (14) --- Black hole physics (159) --- Gravitation (661)}

%% From the front matter, we move on to the body of the paper.
%% Sections are demarcated by \section and \subsection, respectively.
%% Observe the use of the LaTeX \label
%% command after the \subsection to give a symbolic KEY to the
%% subsection for cross-referencing in a \ref command.
%% You can use LaTeX's \ref and \label commands to keep track of
%% cross-references to sections, equations, tables, and figures.
%% That way, if you change the order of any elements, LaTeX will
%% automatically renumber them.
%%
%% We recommend that authors also use the natbib \citep
%% and \citet commands to identify citations.  The citations are
%% tied to the reference list via symbolic KEYs. The KEY corresponds
%% to the KEY in the \bibitem in the reference list below. 

\section{Introduction}
\label{sec:intro}

Nearly all massive galaxies are thought to harbor a massive black hole (MBH) at their centers \citep{kormendy2013coevolution}. When two galaxies merge, dynamical friction is expected to carry their central MBHs from kiloparsec scale separations down to roughly $1-10$ pc \citep[e.g.,][]{mayer2013massive,li2020pairing}. A fraction of these paired MBHs subsequently become gravitationally bound, harden through continued interactions with surrounding stars, gas, and occasionally additional MBHs, and ultimately coalesce as they radiate away energy in gravitational waves \citep[GWs;][]{begelman1980massive}.

Recent pulsar timing array (PTA) detection of a stochastic, low-frequency GW background likely coming from a population of merging massive black hole binaries \citep[MBHBs;][]{agazie2023nanograv_gwbackground, antoniadis2024second, reardon2023search, xu2023searching}, and targeted searches for individual MBHBs \citep[e.g.,][]{agazie2023nanograv,agazie2024nanograv}, underscore the urgent need to understand this population of black holes. In the next decade, the Laser Interferometer Space Antenna \citep[LISA;][]{amaro2017laser, amaro2023astrophysics} is expected to directly observe GWs from individual MBHBs in the $\sim 10^{-4}$ -- $10^{-1}$ Hz band. However, no gravitationally bound or merging MBHBs have yet been conclusively observed.

MBHBs detectable both electromagnetically and with gravitational waves are prime multi-messenger targets: their GW-based distances, combined with electromagnetic (EM) redshifts, provide constraints on the expansion rate of the universe \citep{schutz1986determining,holz2005using}, while measurement of the arrival time difference between GW and EM signals can be used to test general relativity and alternative theories of gravity \citep{kocsis2008premerger,yagi2016black}. These studies highlight the need to identify EM counterparts to MBHBs, and efforts are already underway to search for MBHB candidates using EM signatures \citep{rodriguez2006compact,Eracleous_2012ApJS..201...23E,Graham_2015,Graham_2015Natur.518...74G,Charisi_2016MNRAS.463.2145C,Runnoe_2017MNRAS.468.1683R,Liu_2019,Valtonen_2023,Abdollahi_2024ApJ...976..203A,Rigamonti_2025}.

An MBHB embedded in gas will accrete via a circumbinary disk (CBD) analogous to a single-BH accretion disk. Many simulations have established a general picture of the MBHB--CBD interaction. The binary's gravitational torques carve out a central low-density cavity in the disk, typically with a lopsided eccentric shape. Gas from the inner edge of the cavity is periodically stripped into streams that feed the individual black holes, forming ``minidisks" around each MBH. Some stream material is flung back and collides with the cavity wall. These features---the cavity, streams, minidisks---and their dynamical interactions give rise to distinctive EM variability \citep[see][for a review]{bogdanovic2022electromagnetic}. 

These features have been observed in numerous 2D and 3D simulations of hot and cold\footnote{We refer to cold disks as those with the height-to-radius aspect ratios of $H/R < 0.1$.} disks in hydrodynamical \citep[HD;][]{armitage2005eccentricity,macfadyen2008eccentric,cuadra2009massive,roedig2011limiting,roedig2012evolution,d2013accretion,roedig2014migration,farris2014binary,farris2015binary,farris2015characteristic,nixon2015resonances,young2015binary,ragusa2016suppression,d2016transition,munoz2016pulsed,tang2017orbital,thun2017circumbinary,ryan2017minidisks,miranda2017viscous,tang2018late,munoz2019hydrodynamics,hirsh2020cavity,heath2020orbital,ragusa2020evolution,munoz2020circumbinary,tiede2020gas,duffell2020circumbinary,franchini2021circumbinary,d2021orbital,zrake2021equilibrium,sudarshan2022cooling,smallwood2022accretion,westernacher2022multiband,dittmann2022survey,wang2023role,lai2023circumbinary,franchini2023importance,krauth2023disappearing,dittmann2023decoupling, siwek2023preferential,turpin2024orbital,tiede2024eccentric,franchini2024emission} and in magnetohydrodynamic (MHD) simulations \citep{shi2012three,noble2012circumbinary,farris2012binary,giacomazzo2012general,gold2014accretion,gold2014accretion_two,shi2015three,kelly2017prompt,bowen2017relativistic,bowen2018quasi,bowen2019quasi,paschalidis2021minidisk,armengol2021circumbinary,cattorini2021fully,noble2021mass,combi2022minidisk,avara2023accretion,bright2023minidisc,ruiz2023general,most2024magnetically,Most_2025PhysRevD.111.L081304,Manikantan_2024arXiv241111955M,Manikantan_2025arXiv250412375M,Ennoggi_2025arXiv250910319E,Ennoggi_2025PhRvD.112f3009E}. 

The binary mass ratio, $q = M_2/M_1$, can influence accretion dynamics and observable signatures, where $M_2$ is the mass of the smaller secondary and $M_1$ is the mass of the primary. Near equal-mass binaries ($q \approx 1$) form accretion streams that are roughly symmetric and show accretion surges every half binary orbit. An inner edge overdensity (referred to as the ``lump'') develops in comparable mass binaries, orbiting and inducing quasi-periodic accretion bursts on a timescale of $4-5$ binary orbital periods \citep[see][for a review]{lai2023circumbinary}. Unequal-mass binaries ($q \approx 0.1$), on the other hand, exhibit more asymmetric accretion behavior, where the secondary MBH can accrete disproportionately more than the primary \citep{farris2014binary,young2015binary,duffell2020circumbinary,Dittmann_2021ApJ,siwek2023preferential}. This asymmetry is affected by the CBD's equation of state. \citet{young2015evolution,young2015binary} showed that geometrically thick CBDs can suppress the preferential accretion onto the secondary and, in some cases, favor the primary \citep{ochi2005evolution,hanawa2009gas}. Low mass ratio binaries also don't feature an overdensity at the inner edge \citep{d2013accretion,miranda2017viscous,noble2021mass}. These studies show the impact of mass ratio on binary accretion and, potentially, the observational signatures of MBHBs.

To keep simulations tractable, most models impose simplified gas thermodynamics---usually an isothermal or adiabatic equation of state with approximate heating and cooling schemes. Although this makes the CBD calculations feasible, it also introduces systematic uncertainties in the disk's properties and their observational signatures. These questions can be addressed by radiative magnetohydrodynamics (RMHD) simulations that accurately model the physics of heating and cooling with detailed radiation transport. RMHD simulations of accretion disks around a single black hole, spanning both sub- and super-Eddington regimes, have shown significant changes in disk structure and angular-momentum transport compared to pure MHD simulations \citep[e.g.,][]{jiang2019global,jiang2019super}. Our recent RMHD simulation of a sub-Eddington CBD around an equal-mass MBHB showed that the inclusion of radiation led to a thinner and more filamentary CBD, and the overdensity and streams were weaker compared to a purely MHD case \citep[][hereafter Paper I]{tiwariRMHD}. Similar CBD properties were found in blackbody-cooled disks modeled by \cite{Cocchiararo_2025arXiv250818349C}. Furthermore, our recent RMHD simulation of the minidisks in an equal-mass MBHB shows that the disks become non-axisymmetric, producing anisotropic emission that varies with a half-orbital-period modulation \citep{Chan_2025arXiv250502919C}.

Motivated by the dependence of circumbinary accretion on the mass ratio and building directly on our equal-mass CBD in \citetalias[][]{tiwariRMHD}, here we present an RMHD simulation of a CBD around an unequal-mass MBHB with $q=0.1$. We consider a total binary mass $M_{\rm tot} = M_{1} + M_{2} = 2\times10^7~M_\odot$ on a circular orbit and investigate how radiation affects the structure of the CBD, its accretion rate, and its observational signatures.

The remainder of this paper is organized as follows: in section~\ref{sec:methods}, we describe the simulation setup, in section~\ref{sec:results} we describe the results of our simulations. We discuss the emission properties in section~\ref{sec:disk_em_emission_comp} and their implications in section~\ref{sec:discussion}, and conclude in section~\ref{sec:conclusion}.

\section{Methods}
\label{sec:methods}

\subsection{Binary Configuration}
\label{subsec:binary_config}

Population synthesis studies suggest that MBHBs occupy a broad parameter space, with total masses from $\sim 10^5 - 10^{10} M_{\odot}$, mass ratios of $\sim 10^{-4}-1$, and accretion rates ranging from sub- to super-Eddington \citep{Kelley_2017MNRAS.464.3131K,Katz_2020MNRAS.491.2301K}. Systems with $q \approx 0.1$ represent a substantial fraction of this population. In this work we consider one representative system with mass ratio $q=0.1$ and total mass of $M_{\rm tot} = 2\times10^7\,M_{\odot}$, separated by $a = 100\,r_{\rm g}$ (where $r_{\rm g} = GM_{\rm tot}/c^2$ is the gravitational radius). 

A system with these properties eventually evolves into a GW source for LISA \citep{amaro2017laser,amaro2023astrophysics}. For a binary mass of $2\times10^7\,M_{\odot}$, the separation of $a=100\,r_{\rm g}$ corresponds to $a\approx 9.57 \times 10^{-5}$\,pc, orbital period $t_{\rm orb}$ = $2 \pi (GM_{\rm tot}/a^3)^{-1/2}$ = 7.16 days, and a GW frequency of $f_{\rm GW} = 3.23 \times 10^{-6}$ Hz. Because the time to coalescence due to the emission of GWs \citep[calculated following][]{peters1964gravitational} is $t_{\rm{gw}}$ $\approx$ 73.8 years $\gg t_{\rm orb}$, we neglect orbital evolution due to the emission of GWs and assume that binary orbital separation is fixed over the length of our simulations, which span 100 binary orbits.

The primary and secondary are located at approximately $9\,r_{\rm g}$ and $91\,r_{\rm g}$, respectively, from the center of mass of the system, with Roche lobe radii of about $58\,r_g$ and $21\,r_g$, respectively \citep{eggleton1983approximations}. The inner edge of the CBD resides at approximately $150-200\, r_{\rm g}$. Because the focus of this paper is the CBD, we excise and do not simulate the central region inside $100\, r_{\rm g}$ occupied by the individual mini-disks to avoid issues of the singularities of the individual MBHs. This radius, therefore, marks the inner boundary of the computational domain.

We use a Newtonian potential \citepalias[see][for description]{tiwariRMHD} with the center of mass of the binary as the origin of the spherical polar grid. The MBHB is coplanar with the CBD. The CBD is set up so that its total mass accretion rate is a few times 0.1 Eddington (see Section \ref{subsec:simulation_setup} paragraph 3). Even though the total accretion rate is below the Eddington limit, the lower-mass secondary is expected to accrete at a super-Eddington rate. This is because the secondary, which is closer to the inner edge of the CBD tends to capture a larger fraction of the inflowing gas and, being less massive, has a lower Eddington limit. Therefore, such systems could produce luminous EM emission. The mass of the CBD is much less than the binary, and therefore we neglect the self-gravity of the CBD.

\subsection{Simulation Setup}
\label{subsec:simulation_setup}

Here we summarize the key aspects of the simulation setup and direct the reader to \citetalias{tiwariRMHD} for more details. We solve the ideal MHD equations together with the time-dependent radiative transfer equation for specific intensities as implemented in Athena++ with radiative transfer \citep{stone2020athena++,jiang2014explicit,jiang2021implicit}.

We model the CBD in spherical-polar coordinates $(r,\theta,\phi)$ around a time-varying binary potential \citepalias[equation 2 in][]{tiwariRMHD} and cut out the inner region of the domain at $100\,r_{\rm g}$ containing the two MBHs and their mini-disks. The computational domain extends from $100\,r_{\rm g}$ to $2400\,r_{\rm g}$ in the radial direction ($r$), 0 to $\pi$ in the polar direction ($\theta$), and 0 to $2 \pi$ in the azimuthal direction ($\phi$). We use three levels of static mesh refinement with the finest resolution in the region $r \in(100\, r_{\rm g},500\,r_{\rm g}) \text{, } \theta \in (82.5 \degree,97.5 \degree) \text{, and } \phi \in (0 \degree, 360 \degree)$.
In the polar direction, the grid is progressively refined toward the midplane: $\Delta \theta = 1.875 \degree$ for $\theta \in (0\degree,60 \degree)$, $\Delta \theta = 0.938 \degree$ for $\theta \in (60 \degree,75 \degree)$, $\Delta \theta = 0.469 \degree$ for $\theta \in (75 \degree,82.5 \degree)$, and $\Delta \theta = 0.234 \degree$ in the finest region $\theta \in (82.5 \degree, 90 \degree)$. This refinement structure is symmetrically mirrored across the lower hemisphere. Radially the mesh is logarithmally spaced scaling from $\Delta r = 0.2 r_{\rm g}$ near the inner boundary to $\Delta r=2 r_{\rm g}$ at $500 r_{\rm g}$ and azimuthally $\Delta \phi = 0.469 \degree$ in the highest level of refinement. The effective (maximum) resolution of the grid is $(r \times \theta \times \phi) = (768 \times 768 \times 768)$ and the aspect ratios is $\Delta r/r = \Delta \theta = \Delta \phi/2 = 4.1 \times 10^{-3}$. The boundary conditions are periodic in the $\phi$ direction, polar in $\theta$ and diode in $r$ direction. In the diode boundary conditions, we copy the magnetic and velocity fields of the last active cell in the domain into a ghost cell if the velocity vector is outwards. If the velocity vector points into the simulation domain, we set the radial component of the velocity and magnetic fields to zero while copying other field quantities, along with the pressure and density. This treatment prevents artificial magnetic fields from entering the domain through either boundary.

We initialize the gas with an axisymmetric torus around the binary \citepalias[see][for details]{tiwariRMHD}. The density $\rho(r,\theta)$ and the pressure of the torus $P_{\rm gas}(r,\theta)$ are related by the polytrope: $P_{\rm gas}(r,\theta) = K \rho(r,\theta)^{\eta}$. The torus has a maximum density of $\rho_{m} = 6.17\times10^{-12}$ g\,cm$^{-3}$ at $r_{m} = 600\,r_g$. It is characterized by a polytropic constant $K = 1.09\times10^{8}$ (cm/s)$^2$ and polytropic index $\eta = 1$, with the shear parameter of $s=1.6$. We create a numerical vacuum around the torus with the following properties: $\rho_{\rm m,\text{vac}} = 3.08 \times 10^{-18}$ g cm$^{-3}$, $r_{\rm m,\text{vac}} = 500\, r_{\rm g}$, $\eta_{\text{vac}} = 1$, $K_{\text{vac}}= 1.09 \times 10^{10}\,$ (cm/s)$^2$, and $s_{\text{vac}} = 1.75$. This helps mitigate numerical instabilities that may develop at the edges of the torus.

The magnetic field is defined using the vector potential $\mathbf A(r,\theta) \propto \text{max}(\rho - \frac{1}{2}\rho_{\rm m},0)\ \hat{\mathbf e}_\phi$. This vector potential defines concentric single-loop poloidal magnetic fields that coincide with the $r-\theta$ density contours in the torus. The strength of the fields is set so that initially the volume averaged plasma $\beta$ = $\langle P_{\rm gas} \rangle / \langle P_{\rm mag} \rangle $ = 100. We perturb the pressure by 1\% to seed the magnetorotational instability \citep[MRI;][]{balbus1998instability}.

In order to isolate effects unique to radiation, we conducted two simulations: one MHD with a locally isothermal equation of state (EOS) for the gas and the other as RMHD with adiabatic EOS for the gas with $\gamma = 5/3$. The RMHD simulation is an order of magnitude more computationally expensive than its MHD counterpart. To save computational resources, we start with a single MHD simulation and run it for 40 binary orbits to allow MHD turbulence to develop. At $t=40$ orbits, we branch this shared state into two parallel simulations: one that continues the pure MHD evolution and another that restarts with radiation (RMHD). By doing this, we save about 40 binary orbits worth of computational time in the RMHD simulation.

\begin{figure}
\centering
\includegraphics[width=0.50\textwidth]{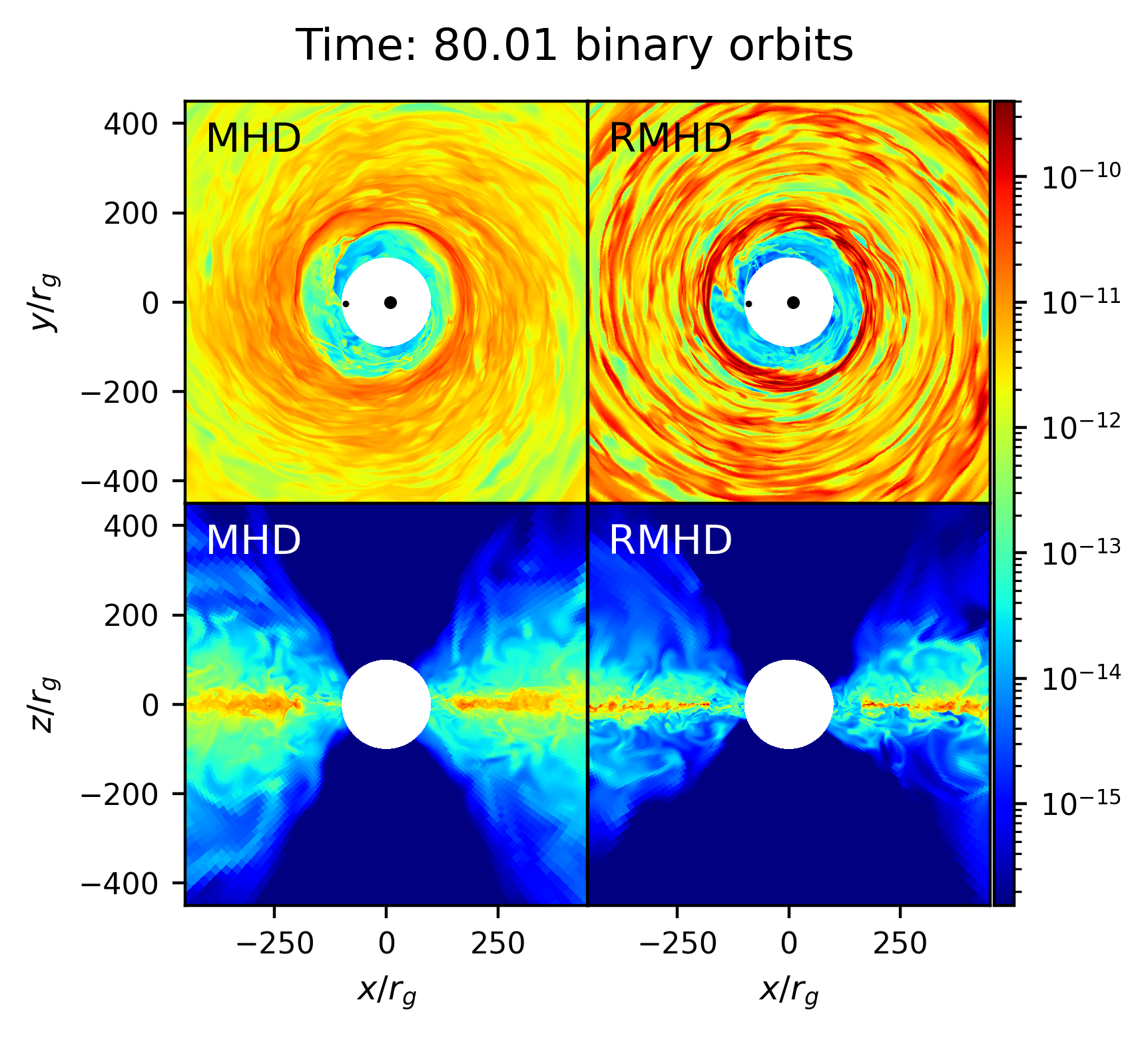}
\caption{\label{fig:density_slices} {\it Top:} Gas density ($\rm g\ cm^{-3}$) in the mid-plane of the disk for the MHD and the RMHD run, showing the denser and more filamentary nature of the disk when radiation is included. {\it Bottom:} Vertical slice showing density ($\rm g\ cm^{-3}$) in the MHD and RMHD runs. The RMHD disk is visibly thinner than the isothermal MHD disk. Both snapshots correspond to a time of approximately 80 binary orbits.}
\end{figure}

In order to incorporate radiation during the restart from the MHD snapshot, we assume that radiation and gas are in local thermodynamic equilibrium (LTE), i.e., $T_{\rm{rad}} = T_{\rm{gas}}$. Moreover, to maintain the same total pressure, we equate the sum of the radiation pressure and gas pressure in RMHD snapshot to the MHD gas pressure ($P_{\rm rad} + P_{\rm{gas,RMHD}} = P_{\rm{gas,MHD}}$). This gives an equation for radiation temperature \citepalias[see equation 10 in][]{tiwariRMHD} that can be numerically solved given the gas pressure from the MHD simulation. Using the radiation temperature, we can set the specific intensities on the angular grid in the lab frame as $I = \frac{\sigma}{\pi} T_{\rm rad}^4$, where $\sigma$ is the Stefan-Boltzmann constant. We use an angular discretization of 80 rays per cell to represent the radiation field. This resolution has been shown to accurately capture radiation anisotropy in the optically thick regions of the flow \citep{davis2014radiation, jiang2014explicit,jiang2019global}. The directions of these 80 rays are distributed on a unit sphere following the discrete ordinates method developed by \citet{1999A&A...348..233B}. The inclusion of radiation following the RMHD restart modifies the effective EOS, initiating a readjustment phase that lasts for approximately 20 binary orbits. To ensure the disk has reached a relaxed state, we conservatively restrict all analysis to $t > 60$ binary orbits.

We adopt Rosseland and Planck mean opacities from the tables of \citet{zhu2021global}, which list opacity values as functions of density and temperature and incorporate atomic, molecular, and dust contributions at solar metallicity.

The MHD simulation covers $\approx$ 100 binary orbits (from $t=0$ to $t=100$ binary orbits), while the RMHD simulation covers 60 binary orbits (from $t=40$ to $t=100$ binary orbits). The fastest-growing MRI mode remains well resolved for the entire MHD run. In contrast, in the RMHD run, the MRI quality factors decline after $\approx$ 80 orbits (corresponding to $\approx$ 10 orbits at $400\,r_{\rm g}$). For this reason, our analysis is limited to the first 80 binary orbits. A detailed discussion of quality factors can be found in Appendix~\ref{subsec:MRI_resolution}.

\begin{figure}
\centering
\includegraphics[trim=0 0 0 0, clip,width=0.40\textwidth]{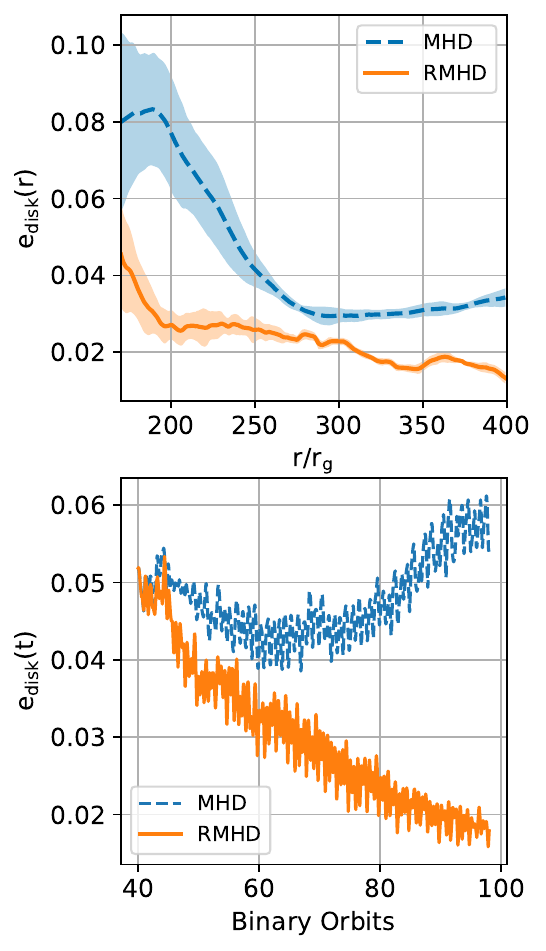}
\caption{\label{fig:disk_eccentricity} {\it Top:} The time and shell-averaged radial disk eccentricity profile for the MHD run (in blue) and the RMHD run (orange). The time averaging was done over the time window of 75-80 binary orbits. Shaded regions show the spread of radial profiles over the time window of 75-80 binary orbits. {\it Bottom:} The time evolution of the inner edge eccentricity in the radial range of $170\,r_g$ to $370\,r_g$ for the MHD (blue) and the RMHD (orange) runs.}
\end{figure}

\begin{figure*}
\centering
\includegraphics[trim=0 0 0 0, clip,width=1\textwidth]{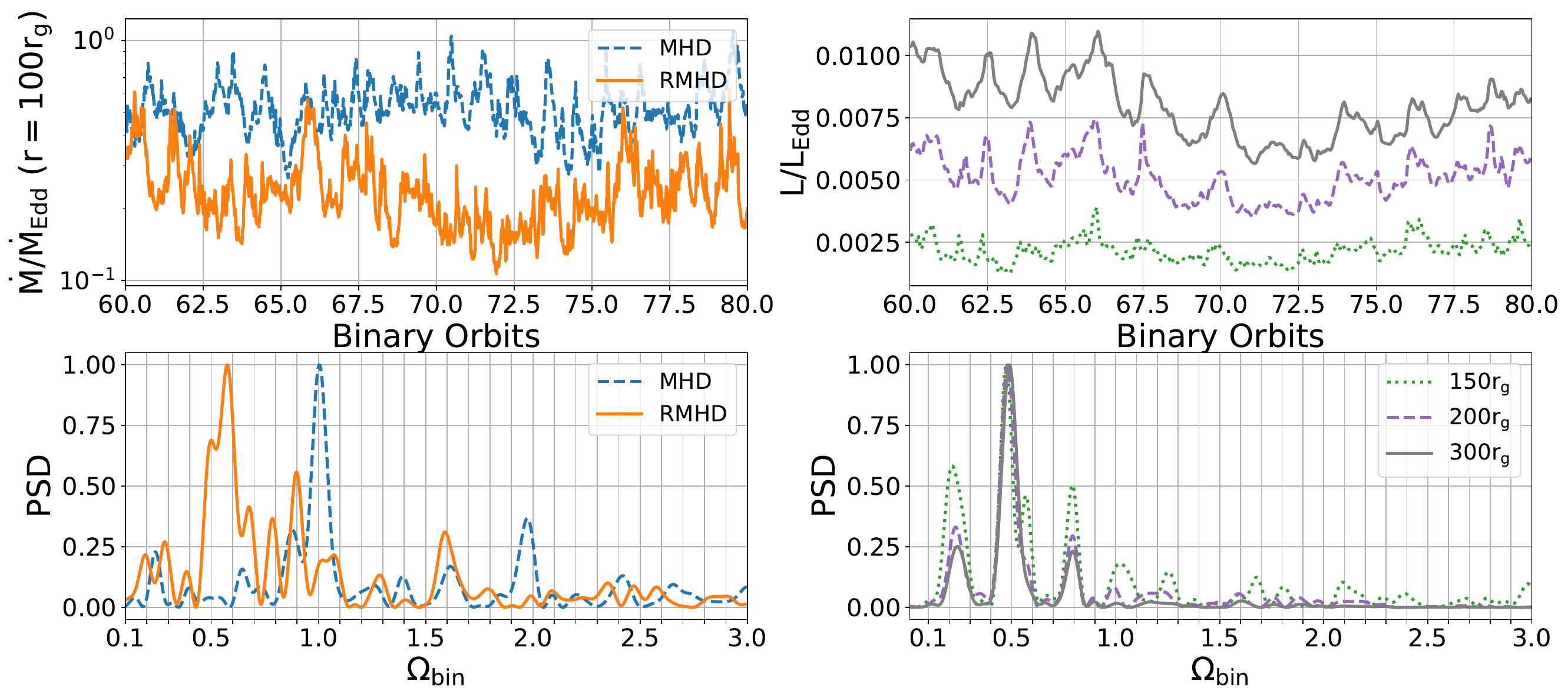}
\caption{\label{fig:Mdot_Lum_total} {\it Top Left:} The total mass accretion rate through the inner edge of the simulation domain at $r=100\,r_{\rm g}$ from 60-80 binary orbits for the MHD run (blue) and the RMHD run (orange). {\it Bottom Left:} The power spectral density of the mass accretion rate for the MHD and the RMHD simulations calculated in the same time window. {\it Top Right:} Luminosity curves calculated on the hemispheres located at $r/r_{\rm g}$ =
150, 200, 300 in the RMHD run. {\it Bottom Right:} Power spectral
density calculated after normalizing the light curves showing a prominent mode at $\approx$ 0.5 $\Omega_{\rm bin}$. The resulting PSDs are normalized to their respective peak values.}
\end{figure*}

\section{Results}
\label{sec:results}

Overall, we find that the RMHD circumbinary disk is thinner, denser, more filamentary, and has a less eccentric inner edge than the MHD disk (Section~\ref{subsec:disk_strcture}).
The total mass accretion rate in the RMHD run is roughly half that in the MHD run and exhibits different quasiperiodic behavior. Accordingly, the associated light curves are also different  (Section~\ref{subsec:mdot_lum}). Finally, radiation in the CBD has little impact on preferential accretion: in both runs, the accretion streams deliver significantly more gas toward the secondary (at roughly twice the rate of the primary's streams), causing accretion towards the secondary to exceed the Eddington limit(Section~\ref{subsec:Preferential Accretion}).

\subsection{Disk Structure}
\label{subsec:disk_strcture}

The top panel of Figure~\ref{fig:density_slices} shows mid‑plane density slices from our MHD and RMHD simulations of the CBD. Binary torques carve out an $\sim170\,r_{\rm g}$ cavity around the binary, with the RMHD disk cavity being more depleted of gas than the MHD disk. Compared to the MHD disk, gas in the RMHD disk is confined into filamentary structures in the midplane slice. These high-density regions are less strongly magnetized. These results are consistent with the results presented in \citetalias{tiwariRMHD} for an equal-mass binary. Furthermore, we note the absence of any significant non-axisymmetric overdensity at the inner edge of the CBD in both MHD and RMHD simulations, consistent with previous simulations of binaries with $q\sim 0.1$ described in the literature \citep{d2013accretion,miranda2017viscous,noble2021mass}.

The bottom panel of Figure~\ref{fig:density_slices} shows the density profile at $\phi=0$, highlighting that the RMHD disk is noticeably thinner and denser in the mid-plane than the MHD disk. This difference arises because radiative cooling causes the disk to lose pressure support and therefore vertical gravity compresses the disk. The same differences were observed in the RMHD simulation of sub-Eddington CBD in \citetalias{tiwariRMHD}. 

Following the methodology of \citetalias{tiwariRMHD}, we determine the density height-to-radius aspect ratio ($H/R$) for both runs. This is calculated by first azimuthally averaging the density, and then identifying the height at a given radius where the density drops by a factor of $e$. This yields an $H/R \sim 0.1$ for the MHD disk and $H/R \sim 0.02$ for the RMHD disk. While our highest refinement boundary at $H/R \sim \pm 0.13$ from the midplane comfortably resolves the thinner RMHD disk, we note that the lower resolution in the regions near the ``wings" of the thicker MHD disk could potentially alter the accretion behavior in those lower-density regions.

The inner edge of the CBD becomes eccentric. The eccentricity can be excited by a combination of streams flung out by the binary that collide with the inner edge of the CBD \citep{shi2012three} and via eccentric Lindblad resonances \citep{lubow1991a,Lubow1991b,lai2023circumbinary}. To understand the effects of radiation on the CBD eccentricity, we compare the radial profiles of eccentricity for the MHD and RMHD runs in Figure \ref{fig:disk_eccentricity} (top panel). We calculate the radial disk profiles of eccentricity as

\begin{equation}
\label{eq:radial_ecc}
e_{disk}(r, t)=\frac{\left|\left\langle\rho v_r e^{i \phi}\right\rangle_{\theta, \phi}\right|}{\left\langle\rho v_\phi\right\rangle_{\theta, \phi}} \;.
\end{equation}
where $\langle X\rangle_{\theta,\phi} = \int X r^2 sin\theta d\theta d\phi / \int r^2 sin\theta d\theta d\phi$ is the shell average of the quantity X. Equation~\ref{eq:radial_ecc} quantifies disk eccentricity by taking the ratio of average fluid element's radial velocity perturbation to its prevailing azimuthal speed, capturing its departure from a perfect circular orbit. Using this approach, we find that the RMHD disk is less eccentric than the MHD disk, with the RMHD disk eccentricity around $\approx 0.04$ while the MHD disk eccentricity $\approx 0.08$ at the inner edge of the disk, and the eccentricity falls with radius in both runs. 

The bottom panel of Figure~\ref{fig:disk_eccentricity} shows the disk eccentricity evolution for the inner CBD, measured over the radial range $r = 170-370\,r_{\rm g}$, corresponding to the region starting at the disk's inner edge ($\sim170\,r_{\rm g}$) and spanning the inner $200\,r_{\rm g}$ of the CBD. The disk eccentricity is calculated as:

\begin{equation}
\label{eq:ecc_time}
e_{disk}(t)=\frac{\left|\int_{170 r_g}^{370 r_g} d r\left\langle\rho v_r e^{i \phi}\right\rangle_{\theta, \phi}\right|}{\int_{170 r_g}^{370 r_g} d r\left\langle\rho v_\phi\right\rangle_{\theta, \phi}} \;.
\end{equation}

Starting from identical values at 40 binary orbits, the MHD disk experiences eccentricity growth while the RMHD disk undergoes eccentricity damping. This difference can be understood in light of the weaker flung-out streams in the RMHD run, which fail to drive the eccentricity evolution at the inner edge.

We observe that radiation suppresses the growth of disk eccentricity compared to the purely MHD run, consistent with longer RHD $\alpha$-disk simulations by \citet{Cocchiararo_2025arXiv250818349C}.

\subsection{Mass Accretion and Luminosity}
\label{subsec:mdot_lum}

MHD stresses caused by the MRI \citep{balbus1998instability} transport angular momentum outwards in the circumbinary disk, causing the infall of gas towards the MBHB. The top left panel of Figure~\ref{fig:Mdot_Lum_total} shows the mass accretion rate across the inner boundary at $r=100\,r_{\rm g}$, which can be taken as the total accretion rate onto both black holes. Over the time window 60-80 binary orbits, the time-averaged accretion rate in the MHD disk is $0.52\,\dot{M}_{\rm Edd}$, which is more than twice that of the RMHD disk at $0.23\,\dot{M}_{\rm Edd}$. Here $\dot{M}_{\rm Edd} = L_{\rm Edd}/\eta c^2$, with $L_{\rm Edd}$ being the Eddington luminosity of a $2\times10^7\,M_{\odot}$ MBH with $\eta = 0.1$. Accretion rate in the RMHD run fluctuates more, with a coefficient of variation (standard deviation/mean) of 0.33 versus 0.23 for MHD.

The power spectral density (PSD) of the mass accretion is shown in the bottom left panel of Figure~\ref{fig:Mdot_Lum_total}. MHD simulation shows a high-frequency mode at one binary period ($2\pi/\Omega_{\rm bin}$), where $\Omega_{\rm bin}$ represents the binary's orbital frequency. This is the consequence of the inner edge of MHD CBD becoming eccentric and the secondary MBH which is closer to the edge drawing gas at the point of closest approach once per orbit.  However, in the RMHD simulation, the power is more broadly distributed across the frequency range $0.5$-$0.9\,\Omega_{\rm bin}$, with a peak near $0.55\,\Omega_{\rm bin}$ (roughly twice the binary orbital period).

We also calculate the corresponding luminosity emitted by the CBD from the RMHD run. Luminosity is obtained by integrating the outward-bound radial flux across three nested hemispheres (we choose $\theta < \pi/2$ hemisphere) at $r/r_{\rm g} = 150, 200, 300$. This accounts for all radiation escaping outward from both optically thick and optically thin regions of the disk while excluding the inwardly advected radiation prevalent in the optically thick regions. Here, we use hemispheres because we are modeling the luminosity as observed by a distant observer viewing the disk from above the mid-plane. The resulting light curves are presented in the top right panel of Figure~\ref{fig:Mdot_Lum_total}. The CBD contributes a total luminosity of $\approx 0.01\,L_{\rm Edd}$ at $300\,r_{\rm g}$.

The power spectral density of the light curves is shown in the bottom right panel of Figure~\ref{fig:Mdot_Lum_total}. At all radii, the luminosity exhibits peak variability at $0.5\,\Omega_{\rm bin}$, associated with twice the binary orbital period. The PSDs of the mass accretion rate and the luminosity show considerably different power distributions, so one should not be used as a proxy for the other. However, the peak power at $0.55\,\Omega_{\rm bin}$ in mass accretion rate roughly matches up with the peak power in the light curves at $0.5\, \Omega_{\rm bin}$. This twice binary orbital signature is in apparent contrast with the $1\,\Omega_{\rm bin}$ peak that dominated our MHD simulation. This indicates that quasi-periodicities in the light curves of MBHBs arise as a combination of orbital motion of the emitting gas and radiation effects, and as a consequence, are non-trivially related to the binary orbital period. 
Moreover, this peak at $0.5\, \Omega_{\rm bin}$ differs from the equal-mass case from \citetalias{tiwariRMHD}, where the dominant periodicities are in the range 0.2-0.4 $\Omega_{\rm bin}$ and $2\, \Omega_{\rm bin}$, indicating some dependence of the mass ratio on the light curves of CBDs.

PSDs for the mass accretion rate and light curves were computed by subtracting running averages to eliminate DC offset, applying a Hann window, and zero-padding the time series prior to performing a Fast Fourier Transform using the scipy.fft function \citep{2020SciPy-NMeth}. The resulting PSDs were normalized to their peak values with frequencies rescaled relative to the binary orbital frequency.

\begin{figure*}
% \label{fig:preferential_accretion}
\centering
\includegraphics[width=1.0\textwidth]{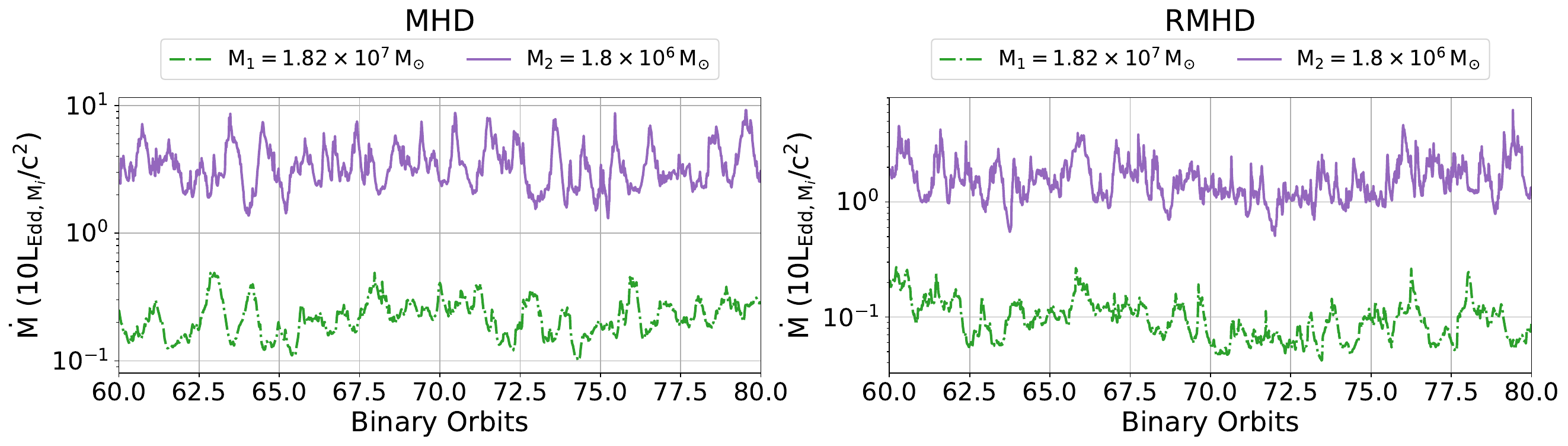}
\caption{\label{fig:superEdd_accretion} {\it Left:} Mass accretion rate through the streams onto the primary (green) and the secondary (purple) scaled by the individual MBHs Eddington limit for the MHD run. {\it Right:} Same as the left panel but for the RMHD run, showing that the secondary is supplied gas at a super-Eddington rate, as in the MHD run.}
\end{figure*}

\subsection{Accretion Streams}
\label{subsec:Preferential Accretion}

Because the secondary MBH orbits closer to the inner edge of the CBD, it accretes more rapidly than the primary, as shown by previous HD simulations \citep{farris2014binary,dittmann2022survey,siwek2023preferential}. A consequence of this preferential accretion is that the secondary can accrete at a super-Eddington rate. Figure~\ref{fig:superEdd_accretion} presents the accretion rates through the streams of the two MBHs calculated at the inner boundary, normalized to each MBH's Eddington limit. The mass accretion rate through the streams $\dot{M}_{1}$ and $\dot{M}_{2}$ onto the primary and secondary are calculated by integrating the inward mass flux ($- \int\rho v_r\, r^2 \sin \theta\, d \theta d \phi$) across two hemispheres at $r = 100\,r_{\rm g}$ that partition the inner domain. The two hemispheres share a plane that passes through the center of mass and is perpendicular to the binary separation axis. Gas crossing the hemisphere closer to the primary MBH is attributed to the primary stream $\dot{M}_1$, and gas crossing the hemisphere closer to the secondary MBH is attributed to the secondary stream $\dot{M}_2$.

Using this approach, we find that in the MHD simulation the primary stream is accreting at $\dot{M}_1 = 5.7\times10^{24}$ g s$^{-1}$ and the secondary stream at $\dot{M}_2 = 8.9\times10^{24}$ g s$^{-1}$ over the time window 60-80 binary orbits. This shows that the primary is supplied gas at a sub-Eddington rate $\dot{M}_1/\dot{M}_{\rm 1,Edd} = 0.23$ whereas the secondary is fed at a super-Eddington rate of $\dot{M}_2/\dot{M}_{\rm 2,Edd} = 3.52$, where $\dot{M}_{\rm 1,Edd}$ and $\dot{M}_{\rm 2,Edd}$ are the Eddington rates defined for the individual MBHs. Similarly in the RMHD simulation, the time-averaged mass supply rate for the primary is $\dot{M}_1 = 2.5\times10^{24}$ g s$^{-1}$ and for the secondary is $\dot{M}_2 = 4.1\times10^{24}$ g s$^{-1}$, showing that the primary is fed at a sub-Eddington rate $\dot{M}_1/\dot{M}_{\rm 1,Edd} = 0.098$ whereas the secondary is fed at a super-Eddington rate of $\dot{M}_2/\dot{M}_{\rm 2,Edd} = 1.61$. In both MHD and RMHD simulation, the inflow rate toward the secondary is super-Eddington, whereas the inflow rate toward the primary remains sub-Eddington.

\begin{figure}
% \label{fig:preferential_accretion}
\centering
\includegraphics[width=0.48\textwidth]{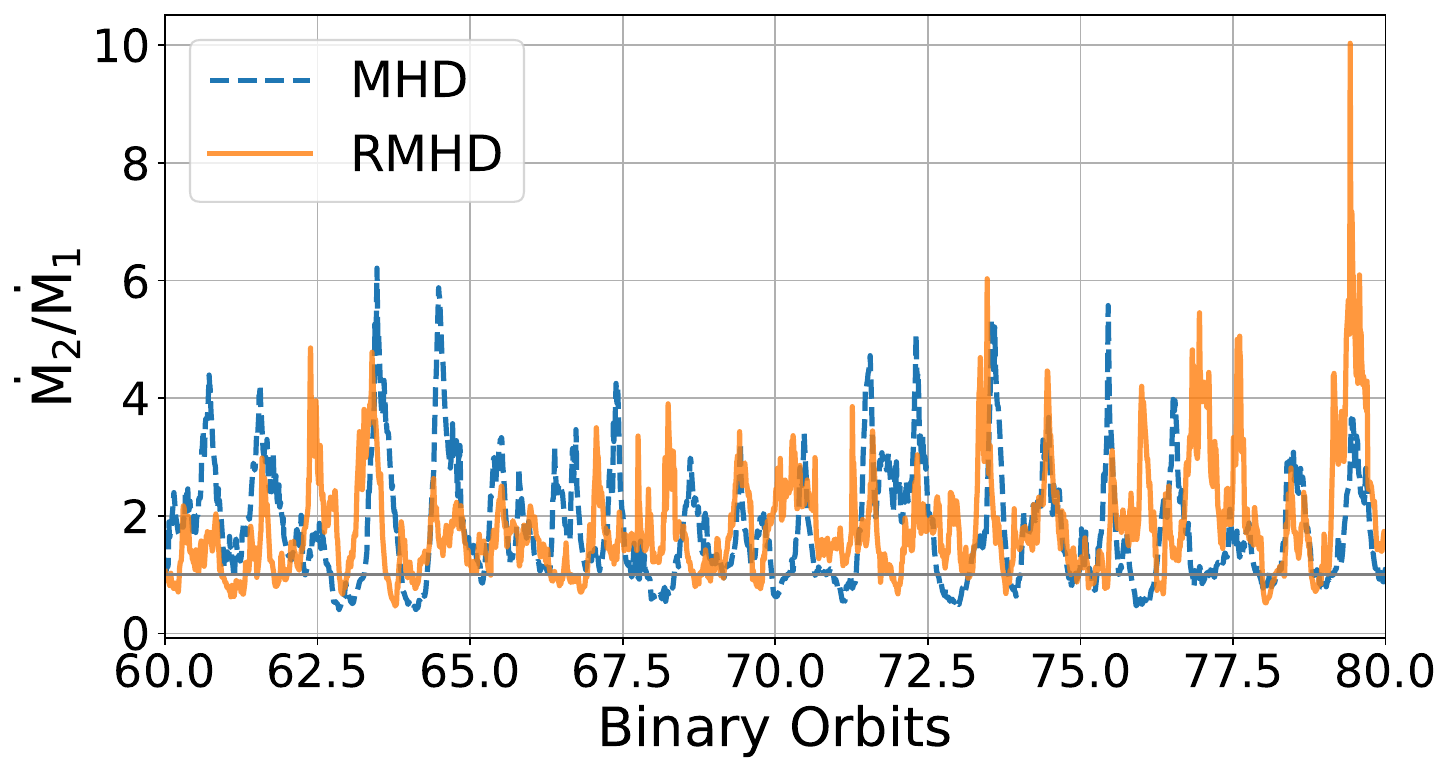}
\caption{\label{fig:preferential_accretion} Ratio of mass accretion rates through the streams onto the MBHs ($\dot{M}_2/\dot{M}_1$) for MHD (blue) and RMHD (blue) simulations. The horizontal gray line marks $\dot{M}_2/\dot{M}_1$=1.}
\end{figure}

Figure~\ref{fig:preferential_accretion} shows the preferential accretion through the streams for the MHD and the RMHD runs, which we quantify as the ratio $\dot{M}_2/\dot{M}_1$. We note that radiation does not impact the ratio $\dot{M}_2/\dot{M}_1$, which averages to 1.76 for MHD and 1.81 for RMHD over the time window 60-80 binary orbits. We note that individual mass accretion rates onto the MBHs can deviate from the mass supplied through the streams due to gas ``sloshing'' between the MBHs via the L1 Lagrange point \citep{avara2023accretion} or the capture of high angular-momentum gas by the companion \citep{tiede2022binaries}. While our work does not resolve these effects, our results show that gas crossing towards the secondary at $100\, r_{\rm g}$ is around twice that of the primary MBH in absolute terms, because of the secondary's proximity to the inner edge of the CBD.

\section{Emission Properties and Comparison with the Equal-Mass Case}
\label{sec:disk_em_emission_comp}

It is interesting to compare our $q=0.1$ RMHD run to the equal-mass RMHD simulation in \citetalias{tiwariRMHD}, which modeled a sub-Eddington CBD around a $2\times10^{7}\,M_{\odot}$ binary separated by $100\,r_{\rm g}$ with the same disk parameters. The equal-mass binary opens a larger central cavity, $\approx 200\,r_{\rm g}$, compared to the lower-mass-ratio system that produces a $\approx 170\,r_{\rm g}$ cavity. The smaller cavity in the $q=0.1$ case results from the weaker binary torques and is consistent with the findings of earlier HD and MHD studies \citep{d2013accretion,shi2015three,ragusa2020evolution,noble2021mass}. Furthermore, the inner edge of the CBD in the equal-mass binary is more eccentric than the $q=0.1$ case. Finally, the equal-mass ratio RMHD simulation led to the formation of an over-density at the inner edge of the CBD, whereas there is no such overdensity in the $q=0.1$ case. This also agrees with earlier HD/MHD CBD simulations \citep{d2013accretion,miranda2017viscous,noble2021mass}. 

To compare the emission properties of our $q=0.1$ run with the equal-mass case, we follow the approach in \citetalias{tiwariRMHD}. First, we locate the Rosseland photosphere - the surface where the optical depth reaches unity, and is computed by integrating:
\begin{equation}
    \tau=\int \rho\, \kappa_r\, r\, d \theta \;,
\end{equation}
from each pole ($\theta =0^{\circ}$ and $\theta = 180^{\circ}$) toward the disk mid-plane. Here, $\kappa_r$ is the Rosseland opacity, $\kappa_{\rm r}= \kappa_{\rm a} + \kappa_{\rm s}$, and, $\kappa_{\rm a}$ and $\kappa_{\rm s}$ are the Rosseland frequency-averaged absorption and scattering opacities, respectively. As demonstrated in \citetalias{tiwariRMHD}, the choice between the Rosseland and effective photospheres, where $\kappa_{\rm eff} = ((\kappa_a+\kappa_s)\kappa_a)^{1/2}$, has a negligible impact on the resulting thermal spectra. We therefore adopt the Rosseland definition for consistency. At the photosphere, we calculate the radiation temperature defined as $T_{\rm rad} = (E_{\rm r}\,c/4 \sigma)^{1/4}$, where $E_{\rm r}$ is the radiation energy density reported by the code, and $\sigma$ is the Stefan-Boltzmann constant. Finally, we calculate the thermal spectrum by integrating the Planck blackbody function over the surface area of the photosphere
\begin{equation}
    L_{\nu} = \int \frac{2 \pi h \nu^3}{c^2}\frac{1}{e^{h \nu/k T_{{\rm rad}}} -1} \, dA \;,
\end{equation}
where $T_{{\rm rad}}(r,\phi) $ is the radiation temperature at the photosphere and $dA = r\, d\phi\, dr$ is the area element. In the calculation of the spectrum, we neglect the emission from $\tau < 1$ regions.

\begin{figure}
\centering
\includegraphics[trim=0 0 0 0, clip,width=0.47\textwidth]{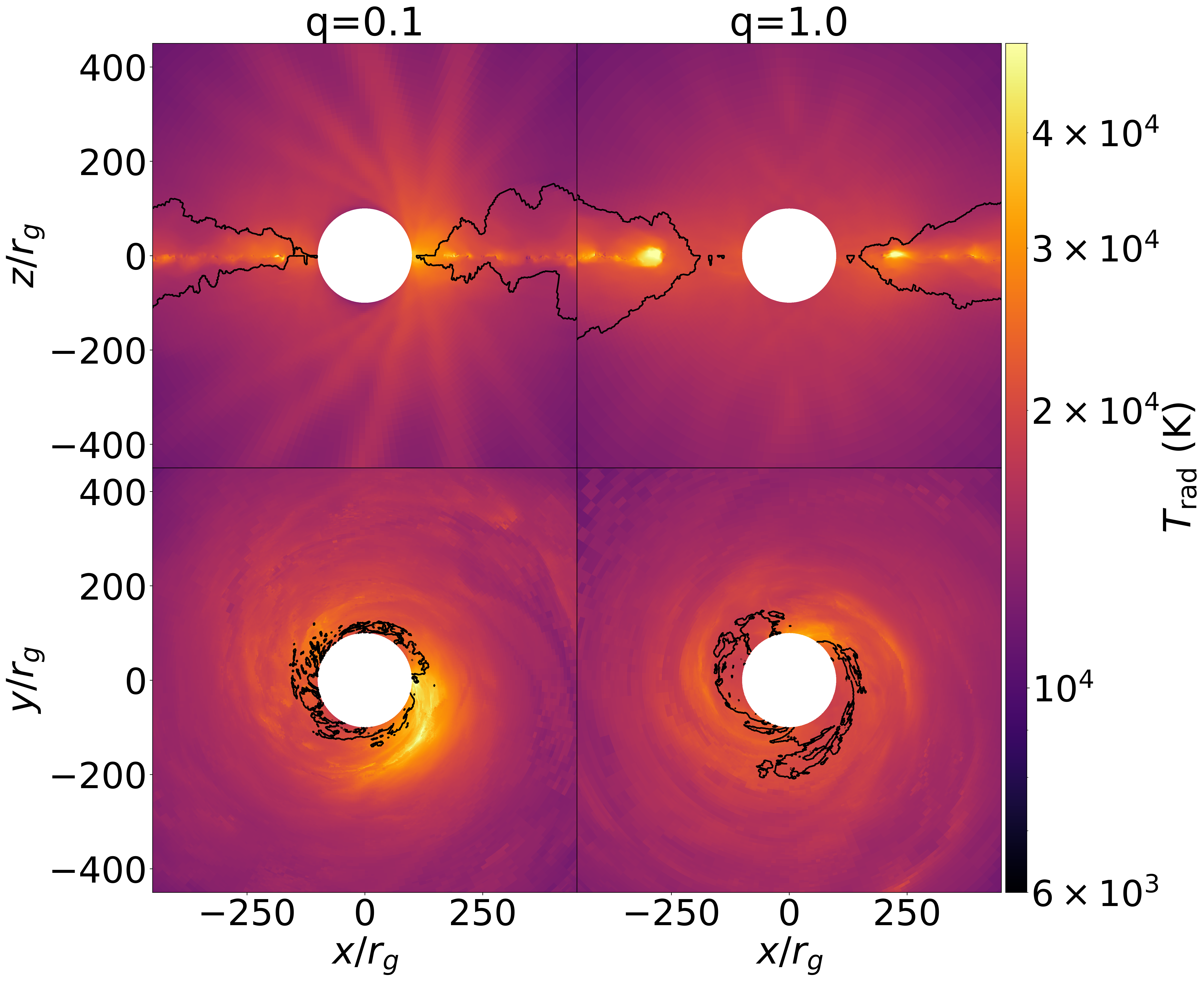}
\caption{\label{fig:rad_temp} Radiation temperature for the $q=0.1$ MBHB at $\approx$ 80 binary orbits (left) and the equal-mass binary from \citetalias[][]{tiwariRMHD} at $\approx$ 73 binary orbits (right). The solid black line shows the photosphere ($\tau = 1$) in both simulations. The top row shows vertical slices and the bottom shows the radiation temperature calculated at the photosphere in the upper hemisphere. In the innermost regions where the photosphere is not defined (because $\tau < 1$), the figure shows the radiation temperature in the mid-plane of the disk.}
\end{figure}

Figure~\ref{fig:rad_temp} compares the radiation temperature of the $q=0.1$ binary (left panels) with the equal-mass binary from \citetalias{tiwariRMHD} (right panels). The top row shows the vertical slices of radiation temperature in the disks, while the bottom row shows the radiation temperature at the photosphere in the upper hemisphere. In both simulations, radiation temperature at the photosphere is on the order of $T_{\rm rad}\sim 10^4\,$K. In the region closest to the central binary, the gas is optically thin ($\tau < 1$), so a photosphere is not defined. Because the $q=0.1$ binary carves out a smaller cavity, its photosphere is located at a smaller average radius than that of the $q=1$ binary. Consequently, the $q=0.1$ photosphere samples comparatively hotter regions, which in turn impacts its spectral signatures, discussed later in the section. %in the next paragraph.

The top panels in Figure~\ref{fig:rad_temp} show artificial rays which appear in the optically thin regions above and below the photosphere. This effect arises from the finite angular discretization of the radiation field in the discrete ordinates method \citep{2019Jrayeffectpaper}, which is here represented by 80 rays emerging from each simulation cell. Because radiation travels in predefined directions, emission can propagate in streaks rather than filling the entire solid angle, particularly in optically thin regions where scattering is negligible. This can ultimately affect the resulting angular distribution of emitted radiation and the properties of the optically thin gas. The radiation energy distribution remains accurate near the photosphere and within the optically thick interior, providing a better description than approximations like the M1 closure or flux-limited diffusion. Furthermore, since our analysis does not depend on angle-resolved radiation properties, and because Athena++ strictly conserves energy, this limitation does not affect the calculation of the radiation luminosity and the resulting bolometric light curves reported in this work.

\begin{figure}
\centering
\includegraphics[trim=0 0 0 0, clip,width=0.5\textwidth]{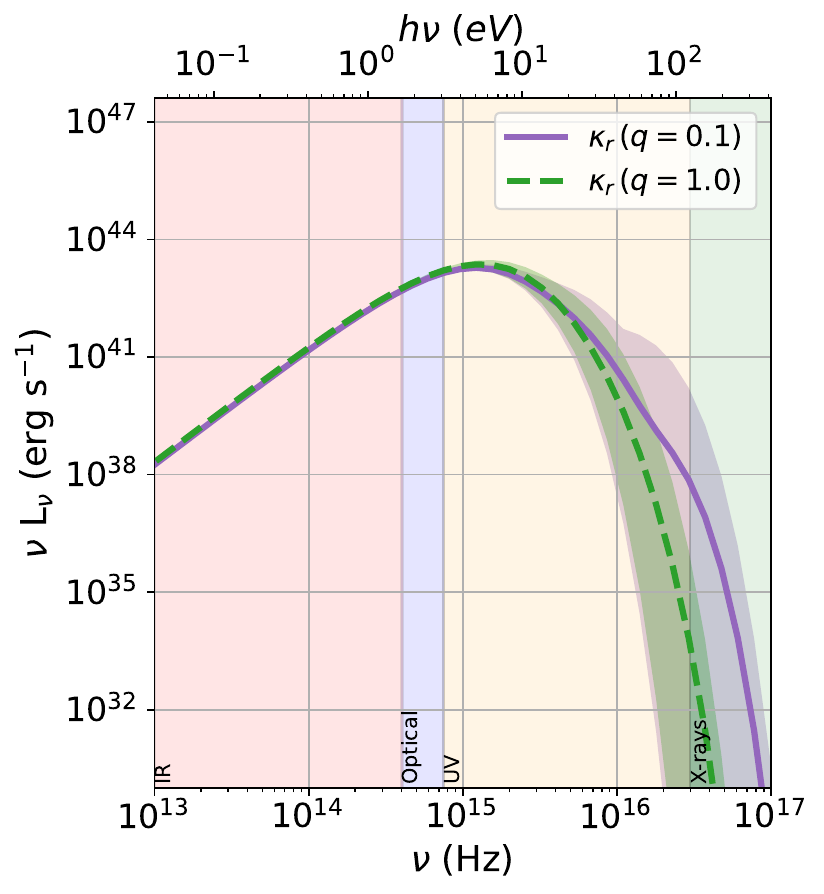}
\caption{\label{fig:spectra} Thermal spectrum calculated from the photosphere for the mass ratio $q=0.1$ (solid purple) and from the equal mass ratio simulation from \citetalias[][]{tiwariRMHD} (dashed green), as described in section~\ref{sec:disk_em_emission_comp}. The vertical shaded regions mark approximate EM bands - infrared (red), optical (blue), UV (yellow), and soft X-ray (green) for visualization purpose.}
\end{figure}

Figure~\ref{fig:spectra} compares thermal spectra from the photosphere of the $q=0.1$ run (solid line) and the equal-mass run (dashed). Both peak in the optical/UV band of the spectrum. Shaded regions show the full spread of spectra over the last ten binary orbits i.e $\approx$ 70-80 orbits for $q=0.1$ and $\approx$ 63 - 73 orbits for $q=1.0$, with the solid curves showing their corresponding means. All curves were computed for regions with $r \le 400\,r_{\rm g}$.  At the high-frequency end (far-UV to soft X-rays) the $q=0.1$ disk is more luminous and variable, fluctuating between $\approx 3\times10^{38}$ -- $1\times10^{42}$ erg s$^{-1}$ at $\sim 30$ eV with a median of $\approx 2\times10^{40}$ erg s$^{-1}$, compared with $\approx 7\times10^{38} - 5\times10^{41}$ erg s$^{-1}$ for the equal-mass case with a median of $\approx 1\times10^{40}$ erg s$^{-1}$. This excess luminosity in $q=0.1$ case originates inside $200\,r_{\rm g}$, where the radiation temperatures are higher and the secondary's closer passage near the photosphere drives stronger fluctuations. This variability in far-UV/soft X-ray might help to distinguish the $q=0.1$ CBD from its equal-mass counterpart. 

For the $q=0.1$ simulation, we also examine the optically thin emission, so we separate the mass accretion into optically thin and thick components, following the method used in \citetalias[][]{tiwariRMHD}. Inside the cavity ($r<170\,r_{\rm g}$) the flow is predominantly optically thin, carrying about 67\% of the total mass accretion rate onto the binary and the remaining 33\% is in optically thick gas. The optically thin material can radiate through thermal bremsstrahlung and synchrotron processes. Using the bremsstrahlung emissivity \citep{rybicki1991radiative}, $\epsilon_{\rm ff} = 1.426 \times 10^{-27} T_{\text {gas }}^{1 / 2}\left(\frac{\rho Z}{\mu m_p}\right)^2 g_{\rm f f}\ \mathrm{erg}\ \mathrm{cm}^{-3}\ \mathrm{sec}^{-1}$ and the local temperature and density within $r\le400\,r_{\rm g}$ in the optically thin regions, we estimate a bremsstrahlung luminosity of $\approx 8\times10^{40}\,{\rm erg\,s^{-1}}$, which is 0.3\% of the thermal luminosity. 

To estimate the synchrotron emission from the CBD, we assume for simplicity that electrons are isothermal with ions. Under this assumption, electrons are non-relativistic. The thermal synchrotron emission \citep{rybicki1991radiative}, $\epsilon_{\rm syn}=\frac{4}{3} \sigma_T c \left(\frac{B^2}{8 \pi}\right)\left(\frac{\rho}{\mu m_p}\right) \frac{\beta_e^2}{1-\beta_e^2}$\ erg\ s$^{-1} \mathrm{~cm}^{-3}$, where $\beta_e^2=\frac{8 k_B T_{\text {gas }}}{\pi m_e c^2}$ assuming a Maxwell-Boltzmann electron distribution, and using the magnetic fields, gas temperatures and gas density from the same optically thin region, the resulting synchrotron luminosity is $\sim9\times10^{41}\,{\rm erg\,s^{-1}}$, or roughly 3\% of the total thermal emission. These luminosities imply that thermal bremsstrahlung will contribute detectable levels of flux in the soft X-ray band ($k_B T_{\rm gas}/h \sim 0.7$ keV), while thermal synchrotron radiation will dominate at radio frequencies ($\nu_{syn} \sim \frac{eB}{\gamma m_e c} \sim 3$ GHz).

\section{Discussion}
\label{sec:discussion}

\subsection{Implications for the Emission Properties of Circumbinary Disks}
\label{subsec:EM_implication}

One key strength of our RMHD simulation is that the location of the photosphere can be calculated self-consistently: gas density and opacity set its position, while the radiation temperature on that surface yields the thermal spectrum from first principles.

For example, for a $2\times10^{7}\,M_{\odot}$ binary separated by $100\,r_{\rm g}$ and accreting at $0.23\,\dot{M}_{\rm Edd}$, our $q=0.1$ run predicts a thermal peak in the near-UV/optical band of the EM spectrum (see section~\ref{sec:disk_em_emission_comp}). Such UV emission is within reach of ULTRASAT \citep{shvartzvald2023ultrasat}, whose 230-290 nm band and minute-cadence monitoring can detect CBDs radiating at $\sim0.01\,L_{\rm Edd}$ to $z\lesssim0.7$ for LISA ($\sim10^{7}\,M_{\odot}$) binaries.

The Vera C. Rubin Observatory \citep{ivezic2019lsst} will be well suited to probe more massive sub-parsec binary systems ($\sim10^{8}-10^{9}\,M_{\odot}$) that are targeted in GWs by the pulsar timing arrays. Assuming $T_{\rm eff}\propto M_{\rm tot}^{-1/4}$, binaries of $\sim10^{8}\,M_{\odot}$ with CBD luminosities of $0.01\,L_{\rm Edd}$ should be visible in the optical to $z\sim1.4$ and display $\sim100$-day modulations - well matched to Rubin's decade-long baseline\footnote{This simple extrapolation ignores possible opacity-driven changes in CBD structure that could modify the spectrum.}. Systems of $\sim10^{9}\,M_{\odot}$ will peak in the optical/near-IR, show $\sim1000$-day modulations, and remain detectable to $z\sim1.0$. Rubin's wide field, high cadence, and long duration thus make it a powerful tool for mapping massive black-hole binaries and their circumbinary environments to $z\sim1$.

\subsection{Limitations of the Calculation}
\label{subsec:cal_limit}

Our RMHD run spans $\approx 100$ binary orbits, but we perform the analysis up to $80$ binary orbits (see Appendix~\ref{subsec:MRI_resolution}). This duration corresponds to about ten orbits at $r=400\,r_{\rm g}$, so we can robustly address phenomena with characteristic timescales shorter than this window. This duration is adequate for studying the inner disk $(r \lesssim 400 r_{\rm g})$ density profile, MRI turbulence, accretion rate, and the emergent luminosity. Calculation of processes that evolve more slowly, such as precession of the disk inner edge or torques acting on the binary, would require much longer and prohibitively expensive simulations. Furthermore, because the viscous inflow timescale for such a thin disk ($H/R \sim 0.02$) is on the order of tens of thousands of binary orbits, the CBD has not reached a true global inflow equilibrium (where the radial mass accretion rate $\dot{M}$ is constant over a wide range of radii). The disk is still undergoing slow secular evolution. However, the relative differences observed between the $q=0.1$ and $q=1.0$ systems provide a robust comparative baseline.

The effects of general relativity are not accounted for. Given the large binary separation ($\gtrsim100\,r_{\rm g}$), relativistic corrections to fluid motion, gravitational-wave back-reaction, and photon trajectories are expected to remain below the percent level. 

We excise the central $100\,r_{\rm g}$, removing the mini-disks and the innermost segments of the accretion streams. Given that the secondary is very likely accreting at a super-Eddington rate, we expect strong radiation-driven outflows and radiation illumination that might affect the inner edge of the CBD. The effects of radiation-driven outflows and radiation illumination on the CBD are not accounted for here and will be studied in future work. Furthermore, by omitting the minidisks, we neglect their X-ray emission, which could potentially mask the high-energy variability we observe from the inner edge of the CBD.

Radiative transfer is handled with gray (frequency-averaged) opacities. Comparison with frequency-dependent RMHD calculations \citep[e.g,][]{mills2024spectral} shows that gray schemes can underestimate gas temperatures in optically thin funnels by failing to treat Compton scattering accurately. Our results are subject to the same limitations that may manifest itself in a lower estimated luminosity of the optically thin region reported in section~\ref{sec:disk_em_emission_comp}. We defer implementation of multi-frequency radiative transport to future work. 

Transient numerical artifacts in the gas pressure occasionally manifest in the RMHD simulation along refinement boundaries. These artifacts occur in the optically thin regions above and below the disk midplane where gas and photons are decoupled. In these regions, gas pressure is typically small compared with the magnetic and radiation pressure, and can be subject to truncation error. Because the gas pressure contributes so little to the total pressure, the numerical artifact in gas pressure does not influence the overall gas dynamics.

Finally, our analysis of emission properties focuses exclusively on the CBD's radiative output. Two observational caveats are not modeled: (i) the host galaxy may outshine the CBD emission, especially in low-mass hosts typical of LISA sources, and (ii) UV photons can be absorbed by interstellar or intergalactic material. Consequently, the luminosities and spectra reported here should be regarded as upper limits.

\section{Conclusions}
\label{sec:conclusion}

We carry out a global RMHD simulation of a sub-Eddington circumbinary disk around a $2 \times 10^7 M_{\odot}$ binary with mass ratio $q=0.1$ and orbital separation of $100\,r_{\rm g}$, representative of a source that eventually evolves into the LISA GW band. We examine the disk's structure, accretion properties and emission signatures and compare them to those simulated with MHD. We also compare the emission and disk properties with the RMHD simulation of the equal-mass ratio binary from \citetalias{tiwariRMHD}. Our main findings are:

\begin{itemize}
    \item There are substantial differences in the structure of the CBD in the MHD and RMHD runs. Similar to \citetalias[][]{tiwariRMHD}, we find that the CBD simulated with RMHD is thinner, denser, and more filamentary than its MHD counterpart. The MHD disk is more lopsided and eccentric than the RMHD disk.
    \item The inclusion of radiation reduces the total mass accretion rate (0.23 $\dot{M}_{\rm Edd}$ for the RMHD vs 0.52 $\dot{M}_{\rm Edd}$ in MHD) but doesn't seem to affect preferential accretion onto the secondary MBH ($\dot{M}_2/\dot{M}_1$), which is close to 1.8 in both runs.
    \item The mass accretion in the RMHD simulation powers the thermal emission from the inner regions of the CBD at 1\% $L_{\rm Edd}$, which for a $2\times10^{7}\,M_{\odot}$ MBHB peaks in the optical/UV band. In the RMHD run, the PSDs of the mass accretion rate and the luminosity show considerably different power distributions, so one should not be used as a proxy for the other. The calculated light curve from the RMHD run shows quasi-periodicity at approximately $0.5\, \Omega_{\rm bin}$ for the $q=0.1$ binary, whereas the equal-mass binary from \citetalias{tiwariRMHD} showed dominant periodicities in the range 0.2-0.4 $\Omega_{\rm bin}$ and $2\, \Omega_{\rm bin}$.
    \item The photosphere in the $q=0.1$ RMHD simulation reaches deeper into the CBD cavity as compared to the equal-mass binary, where it samples the hot turbulent gas, perturbed by the secondary MBH. As a result the thermal spectrum of the $q=0.1$ MBHB system is more luminous and shows higher variability in the far-UV and soft X-rays compared to the equal-mass binary. For example, near $h\nu \approx 30$ eV the $q=0.1$ system is roughly a factor of two brighter ($2\times10^{40}$ erg s$^{-1}$ vs $1\times10^{40}$ erg s$^{-1}$) and varies by a factor of 3000 ($3\times\sim 10^{38} - 10^{42}$ erg s$^{-1}$), compared to a factor of $\sim$ 700 for the equal-mass binary ($\sim 7\times10^{38} - 5\times10^{41}$ erg s$^{-1}$).
    \item In our RMHD CBD, a substantial fraction of the cavity gas is optically thin, making it a significant source of thermal synchrotron emission in the radio band and to a smaller degree, thermal bremsstrahlung emission in the soft X-ray band of the spectrum, providing additional avenues for the CBD detection.
\end{itemize}

This study is a simulation of a circumbinary accretion disk with self-consistent thermodynamics that couples radiative transfer with the dynamical evolution of gas to predict time-dependent EM counterparts. Together with \citetalias{tiwariRMHD}, this work is a stepping stone towards the future end-to-end models that track binaries from inspiral to merger, therefore directly linking theory with observations. In these models, once the binary parameters are fixed, the gas accretion rate is the sole free parameter required to predict the full set of EM signatures from an MBHB. That, in turn, will allow us to build a template bank of EM counterparts, much like existing GW waveform catalogs. With sustained effort, the field could achieve this within the next five to ten years - just in time for the first individual PTA detections and before LISA's launch.

\begin{acknowledgments}
The authors are grateful to the anonymous referee for constructive comments that improved the manuscript. This work was supported by the National Aeronautics and Space Administration (NASA) under grant 80NSSC19K0319, by the National Science Foundation (NSF) under grant AST-1908042, and by the Research Corporation for Science Advancement under award CS-SEED-2023-008. Research cyberinfrastructure resources and services supporting this work were provided in part by the NASA High-End Computing (HEC) Program through the NASA Advanced Supercomputing (NAS) Division at Ames Research Center, and in part by the Partnership for an Advanced Computing Environment (PACE) at the Georgia Institute of Technology, Atlanta, Georgia, United States of America. 
The Center for Computational Astrophysics at the
Flatiron Institute is supported by the Simons Foundation.
\end{acknowledgments}

\vspace{5mm}

\software{Athena++ \citep{stone2020athena++,jiang2021implicit},  
          Matplotlib \citep{Hunter:2007}, 
          Numpy \citep{harris2020array},
          mpi4py \citep{dalcin2005mpi}
          }

\twocolumngrid

\appendix

\section{Resolving the MRI}
\label{subsec:MRI_resolution}

\begin{figure}[ht!]
    \centering
    \includegraphics[width=0.90\textwidth]{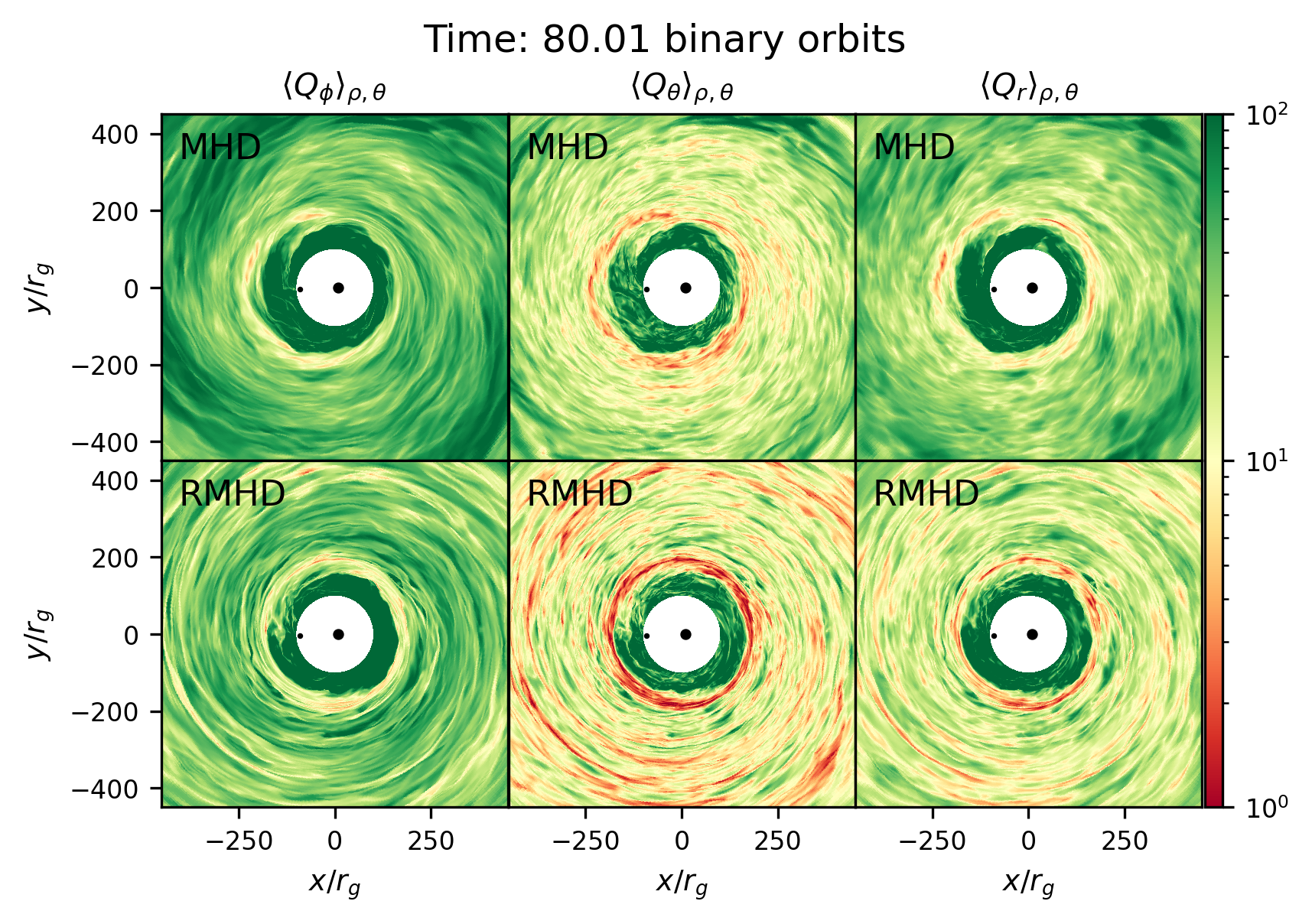}
    \caption{Density weighted vertically averaged quality factors measured at $t \approx 80$ binary orbits for the MHD (top) and RMHD (bottom) run: $Q_{\phi}$ (left), $Q_{\theta}$ (middle), $Q_{r}$ (right).}
    \label{fig:vertical_avg_quality_factors}
\end{figure}

The magnetorotational instability \citep{balbus1998instability} drives correlated MHD turbulence, and the attendant MHD stresses are responsible for driving accretion. The need to resolve the nonlinear development of the magnetorotational instability sets a strict requirement on the numerical resolution in simulations of magnetized accretion flows. We test our grid against this standard by computing the characteristic MRI wavelength, $\lambda_{i} = (2 \pi \sqrt{16/15} |v_{A,i}| / \Omega)$, where $|v_{A,i}|$ is the Alfv\'en velocity in $i = \phi, \theta$ or $r$ directions and use it to evaluate the corresponding quality factors. The quality factors compare the characteristic MRI wavelength to the local grid spacing, $Q_{i} = \lambda_{i} / \Delta x_{i}$, where $\Delta x_{i}$ is $r \text{sin} \theta \Delta \phi$ along the azimuthal direction, $r \Delta \theta$ along the polar direction, and $\Delta r$ along the radial direction. Figure \ref{fig:vertical_avg_quality_factors} shows the density-weighted vertical average for $Q_{\phi}, Q_{\theta}, Q_{r}$, for the MHD and RMHD runs.

We apply the MHD resolution criteria originally derived for non-radiative disks to both our MHD and RMHD runs. MRI turbulence is considered adequately resolved when $Q_{\phi} \gtrsim 25$ and $Q_{\theta} \gtrsim 10$ \citep{sorathia2012global,hawley2013testing}. Throughout all of the MHD disk, these thresholds are met or exceeded until the end of the simulation at around $\approx 100$ binary orbits. In the RMHD disk after $\approx$ 80 binary orbits the above criteria are no longer met and as a result, we analyze the results only up to 80 binary orbits. We also note that lower $Q$ values occur where the gas is dense and magnetic fields are weak, notably near the disk's inner edge and within the RMHD filaments, similar to the findings noted by \cite{noble2021mass} and in \citetalias{tiwariRMHD}. Although reduced quality factors at the inner edge of the cavity slows MRI growth there, it has little effect on our results because gravitational torques, not MRI stresses, dominate angular-momentum transport and mass inflow in that region.

%%%%%%%%%%%%%%%%%%%%%%%%%%%%%%%%%%%%%%%%%%%%%%%%%
%%%  FIGURE 19
%%%%%%%%%%%%%%%%%%%%%%%%%%%%%%%%%%%%%%%%%%%%%%%%%

\bibliography{sample631}{}

@ARTICLE{li2020pairing,
       author = {{Li}, Kunyang and {Bogdanovi{\'c}}, Tamara and {Ballantyne}, David R.},
        title = "{Pairing of Massive Black Holes in Merger Galaxies Driven by Dynamical Friction}",
      journal = {\apj},
         year = 2020,
        month = jun,
       volume = {896},
       number = {2},
          eid = {113},
        pages = {113},
          doi = {10.3847/1538-4357/ab93c6},
archivePrefix = {arXiv},
       eprint = {2006.08520},
 primaryClass = {astro-ph.GA},
       adsurl = {https://ui.adsabs.harvard.edu/abs/2020ApJ...896..113L}
}

@ARTICLE{begelman1980massive,
       author = {{Begelman}, M.~C. and {Blandford}, R.~D. and {Rees}, M.~J.},
        title = "{Massive black hole binaries in active galactic nuclei}",
      journal = {\nat},
         year = 1980,
        month = sep,
       volume = {287},
       number = {5780},
        pages = {307-309},
          doi = {10.1038/287307a0},
       adsurl = {https://ui.adsabs.harvard.edu/abs/1980Natur.287..307B}
}

@ARTICLE{stone2020athena++,
       author = {{Stone}, James M. and {Tomida}, Kengo and {White}, Christopher J. and {Felker}, Kyle G.},
        title = "{The Athena++ Adaptive Mesh Refinement Framework: Design and Magnetohydrodynamic Solvers}",
      journal = {\apjs},
         year = 2020,
        month = jul,
       volume = {249},
       number = {1},
          eid = {4},
        pages = {4},
          doi = {10.3847/1538-4365/ab929b},
archivePrefix = {arXiv},
       eprint = {2005.06651},
 primaryClass = {astro-ph.IM},
       adsurl = {https://ui.adsabs.harvard.edu/abs/2020ApJS..249....4S}
}

@ARTICLE{tiede2020gas,
       author = {{Tiede}, Christopher and {Zrake}, Jonathan and {MacFadyen}, Andrew and {Haiman}, Zoltan},
        title = "{Gas-driven Inspiral of Binaries in Thin Accretion Disks}",
      journal = {\apj},
         year = 2020,
        month = sep,
       volume = {900},
       number = {1},
          eid = {43},
        pages = {43},
          doi = {10.3847/1538-4357/aba432},
archivePrefix = {arXiv},
       eprint = {2005.09555},
 primaryClass = {astro-ph.GA},
       adsurl = {https://ui.adsabs.harvard.edu/abs/2020ApJ...900...43T}
}

@ARTICLE{mayer2013massive,
       author = {{Mayer}, Lucio},
        title = "{Massive black hole binaries in gas-rich galaxy mergers; multiple regimes of orbital decay and interplay with gas inflows}",
      journal = {Classical and Quantum Gravity},
         year = 2013,
        month = dec,
       volume = {30},
       number = {24},
          eid = {244008},
        pages = {244008},
          doi = {10.1088/0264-9381/30/24/244008},
archivePrefix = {arXiv},
       eprint = {1308.0431},
 primaryClass = {astro-ph.CO},
       adsurl = {https://ui.adsabs.harvard.edu/abs/2013CQGra..30x4008M}
}

@article{amaro2017laser,
  title={Laser interferometer space antenna},
  author={Amaro-Seoane, Pau and Audley, Heather and Babak, Stanislav and Baker, John and Barausse, Enrico and Bender, Peter and Berti, Emanuele and Binetruy, Pierre and Born, Michael and Bortoluzzi, Daniele and others},
  journal={arXiv preprint arXiv:1702.00786},
  year={2017}
}

@ARTICLE{kormendy2013coevolution,
       author = {{Kormendy}, John and {Ho}, Luis C.},
        title = "{Coevolution (Or Not) of Supermassive Black Holes and Host Galaxies}",
      journal = {\araa},
         year = 2013,
        month = aug,
       volume = {51},
       number = {1},
        pages = {511-653},
          doi = {10.1146/annurev-astro-082708-101811},
archivePrefix = {arXiv},
       eprint = {1304.7762},
 primaryClass = {astro-ph.CO},
       adsurl = {https://ui.adsabs.harvard.edu/abs/2013ARA&A..51..511K}
}

@article{schutz1986determining,
  title={Determining the Hubble constant from gravitational wave observations},
  author={Schutz, Bernard F},
  journal={Nature},
  volume={323},
  number={6086},
  pages={310--311},
  year={1986},
  publisher={Nature Publishing Group}
}

@article{holz2005using,
  title={Using gravitational-wave standard sirens},
  author={Holz, Daniel E and Hughes, Scott A},
  journal={The Astrophysical Journal},
  volume={629},
  number={1},
  pages={15},
  year={2005},
  publisher={IOP Publishing}
}

@article{kocsis2008premerger,
  title={Premerger localization of gravitational wave standard sirens with LISA: triggered search for an electromagnetic counterpart},
  author={Kocsis, Bence and Haiman, Zoltan and Menou, Kristen},
  journal={The Astrophysical Journal},
  volume={684},
  number={2},
  pages={870},
  year={2008},
  publisher={IOP Publishing}
}

@article{yagi2016black,
  title={Black hole based tests of general relativity},
  author={Yagi, Kent and Stein, Leo C},
  journal={Classical and Quantum Gravity},
  volume={33},
  number={5},
  pages={054001},
  year={2016},
  publisher={IOP Publishing}
}

@article{balbus1998instability,
  title={Instability, turbulence, and enhanced transport in accretion disks},
  author={Balbus, Steven A and Hawley, John F},
  journal={Reviews of modern physics},
  volume={70},
  number={1},
  pages={1},
  doi={10.1103/RevModPhys.70.1},
  year={1998},
  publisher={APS}
}

@article{armitage2005eccentricity,
  title={Eccentricity of supermassive black hole binaries coalescing from gas-rich mergers},
  author={Armitage, Philip J and Natarajan, Priyamvada},
  journal={The Astrophysical Journal},
  volume={634},
  number={2},
  pages={921},
  doi={10.1086/497108},
  year={2005},
  publisher={IOP Publishing}
}

@article{macfadyen2008eccentric,
  title={An eccentric circumbinary accretion disk and the detection of binary massive black holes},
  author={MacFadyen, Andrew I and Milosavljevi{\'c}, Milo{\v{s}}},
  journal={The Astrophysical Journal},
  volume={672},
  number={1},
  pages={83},
  doi={10.1086/523869},
  year={2008},
  publisher={IOP Publishing}
}

@ARTICLE{cuadra2009massive,
       author = {{Cuadra}, J. and {Armitage}, P.~J. and {Alexander}, R.~D. and {Begelman}, M.~C.},
        title = "{Massive black hole binary mergers within subparsec scale gas discs}",
      journal = {\mnras},
         year = 2009,
        month = mar,
       volume = {393},
       number = {4},
        pages = {1423-1432},
          doi = {10.1111/j.1365-2966.2008.14147.x},
archivePrefix = {arXiv},
       eprint = {0809.0311},
 primaryClass = {astro-ph},
       adsurl = {https://ui.adsabs.harvard.edu/abs/2009MNRAS.393.1423C}
}

@ARTICLE{shi2012three,
       author = {{Shi}, Ji-Ming and {Krolik}, Julian H. and {Lubow}, Stephen H. and {Hawley}, John F.},
        title = "{Three-dimensional Magnetohydrodynamic Simulations of Circumbinary Accretion Disks: Disk Structures and Angular Momentum Transport}",
      journal = {\apj},
         year = 2012,
        month = apr,
       volume = {749},
       number = {2},
          eid = {118},
        pages = {118},
          doi = {10.1088/0004-637X/749/2/118},
archivePrefix = {arXiv},
       eprint = {1110.4866},
 primaryClass = {astro-ph.HE},
       adsurl = {https://ui.adsabs.harvard.edu/abs/2012ApJ...749..118S}
}

@ARTICLE{shi2015three,
       author = {{Shi}, Ji-Ming and {Krolik}, Julian H.},
        title = "{Three-dimensional MHD Simulation of Circumbinary Accretion Disks. II. Net Accretion Rate}",
      journal = {\apj},
         year = 2015,
        month = jul,
       volume = {807},
       number = {2},
          eid = {131},
        pages = {131},
          doi = {10.1088/0004-637X/807/2/131},
archivePrefix = {arXiv},
       eprint = {1503.05561},
 primaryClass = {astro-ph.HE},
       adsurl = {https://ui.adsabs.harvard.edu/abs/2015ApJ...807..131S}
}

@ARTICLE{noble2012circumbinary,
       author = {{Noble}, Scott C. and {Mundim}, Bruno C. and {Nakano}, Hiroyuki and {Krolik}, Julian H. and {Campanelli}, Manuela and {Zlochower}, Yosef and {Yunes}, Nicol{\'a}s},
        title = "{Circumbinary Magnetohydrodynamic Accretion into Inspiraling Binary Black Holes}",
      journal = {\apj},
         year = 2012,
        month = aug,
       volume = {755},
       number = {1},
          eid = {51},
        pages = {51},
          doi = {10.1088/0004-637X/755/1/51},
archivePrefix = {arXiv},
       eprint = {1204.1073},
 primaryClass = {astro-ph.HE},
       adsurl = {https://ui.adsabs.harvard.edu/abs/2012ApJ...755...51N}
}

@ARTICLE{d2013accretion,
       author = {{D'Orazio}, Daniel J. and {Haiman}, Zolt{\'a}n and {MacFadyen}, Andrew},
        title = "{Accretion into the central cavity of a circumbinary disc}",
      journal = {\mnras},
         year = 2013,
        month = dec,
       volume = {436},
       number = {4},
        pages = {2997-3020},
          doi = {10.1093/mnras/stt1787},
archivePrefix = {arXiv},
       eprint = {1210.0536},
 primaryClass = {astro-ph.GA},
       adsurl = {https://ui.adsabs.harvard.edu/abs/2013MNRAS.436.2997D}
}

@ARTICLE{farris2014binary,
       author = {{Farris}, Brian D. and {Duffell}, Paul and {MacFadyen}, Andrew I. and {Haiman}, Zoltan},
        title = "{Binary Black Hole Accretion from a Circumbinary Disk: Gas Dynamics inside the Central Cavity}",
      journal = {\apj},
         year = 2014,
        month = mar,
       volume = {783},
       number = {2},
          eid = {134},
        pages = {134},
          doi = {10.1088/0004-637X/783/2/134},
archivePrefix = {arXiv},
       eprint = {1310.0492},
 primaryClass = {astro-ph.HE},
       adsurl = {https://ui.adsabs.harvard.edu/abs/2014ApJ...783..134F}
}

@ARTICLE{tang2017orbital,
       author = {{Tang}, Yike and {MacFadyen}, Andrew and {Haiman}, Zolt{\'a}n},
        title = "{On the orbital evolution of supermassive black hole binaries with circumbinary accretion discs}",
      journal = {\mnras},
         year = 2017,
        month = aug,
       volume = {469},
       number = {4},
        pages = {4258-4267},
          doi = {10.1093/mnras/stx1130},
archivePrefix = {arXiv},
       eprint = {1703.03913},
 primaryClass = {astro-ph.HE},
       adsurl = {https://ui.adsabs.harvard.edu/abs/2017MNRAS.469.4258T}
}

@ARTICLE{miranda2017viscous,
       author = {{Miranda}, Ryan and {Mu{\~n}oz}, Diego J. and {Lai}, Dong},
        title = "{Viscous hydrodynamics simulations of circumbinary accretion discs: variability, quasi-steady state and angular momentum transfer}",
      journal = {\mnras},
         year = 2017,
        month = apr,
       volume = {466},
       number = {1},
        pages = {1170-1191},
          doi = {10.1093/mnras/stw3189},
archivePrefix = {arXiv},
       eprint = {1610.07263},
 primaryClass = {astro-ph.SR},
       adsurl = {https://ui.adsabs.harvard.edu/abs/2017MNRAS.466.1170M}
}

@ARTICLE{munoz2019hydrodynamics,
       author = {{Mu{\~n}oz}, Diego J. and {Miranda}, Ryan and {Lai}, Dong},
        title = "{Hydrodynamics of Circumbinary Accretion: Angular Momentum Transfer and Binary Orbital Evolution}",
      journal = {\apj},
         year = 2019,
        month = jan,
       volume = {871},
       number = {1},
          eid = {84},
        pages = {84},
          doi = {10.3847/1538-4357/aaf867},
archivePrefix = {arXiv},
       eprint = {1810.04676},
 primaryClass = {astro-ph.HE},
       adsurl = {https://ui.adsabs.harvard.edu/abs/2019ApJ...871...84M}
}

@ARTICLE{dittmann2022survey,
       author = {{Dittmann}, Alexander J. and {Ryan}, Geoffrey},
        title = "{A survey of disc thickness and viscosity in circumbinary accretion: Binary evolution, variability, and disc morphology}",
      journal = {\mnras},
         year = 2022,
        month = jul,
       volume = {513},
       number = {4},
        pages = {6158-6176},
          doi = {10.1093/mnras/stac935},
archivePrefix = {arXiv},
       eprint = {2201.07816},
 primaryClass = {astro-ph.HE},
       adsurl = {https://ui.adsabs.harvard.edu/abs/2022MNRAS.513.6158D}
}

@article{rodriguez2006compact,
  title={A compact supermassive binary black hole system},
  author={Rodriguez, C and Taylor, Greg B and Zavala, RT and Peck, AB and Pollack, LK and Romani, RW},
  journal={The Astrophysical Journal},
  volume={646},
  number={1},
  pages={49},
  year={2006},
  publisher={IOP Publishing}
}

@ARTICLE{bogdanovic2022electromagnetic,
       author = {{Bogdanovi{\'c}}, Tamara and {Miller}, M. Coleman and {Blecha}, Laura},
        title = "{Electromagnetic counterparts to massive black-hole mergers}",
      journal = {Living Reviews in Relativity},
         year = 2022,
        month = dec,
       volume = {25},
       number = {1},
          eid = {3},
        pages = {3},
          doi = {10.1007/s41114-022-00037-8},
archivePrefix = {arXiv},
       eprint = {2109.03262},
 primaryClass = {astro-ph.HE},
       adsurl = {https://ui.adsabs.harvard.edu/abs/2022LRR....25....3B}
}

@ARTICLE{jiang2019global,
       author = {{Jiang}, Yan-Fei and {Blaes}, Omer and {Stone}, James M. and {Davis}, Shane W.},
        title = "{Global Radiation Magnetohydrodynamic Simulations of sub-Eddington Accretion Disks around Supermassive Black Holes}",
      journal = {\apj},
         year = 2019,
        month = nov,
       volume = {885},
       number = {2},
          eid = {144},
        pages = {144},
          doi = {10.3847/1538-4357/ab4a00},
archivePrefix = {arXiv},
       eprint = {1904.01674},
 primaryClass = {astro-ph.HE},
       adsurl = {https://ui.adsabs.harvard.edu/abs/2019ApJ...885..144J}
}

@ARTICLE{peters1964gravitational,
       author = {{Peters}, P.~C.},
        title = "{Gravitational Radiation and the Motion of Two Point Masses}",
      journal = {Physical Review},
         year = 1964,
        month = nov,
       volume = {136},
       number = {4B},
        pages = {1224-1232},
          doi = {10.1103/PhysRev.136.B1224},
       adsurl = {https://ui.adsabs.harvard.edu/abs/1964PhRv..136.1224P}
}

@ARTICLE{eggleton1983approximations,
       author = {{Eggleton}, P.~P.},
        title = "{Aproximations to the radii of Roche lobes.}",
      journal = {\apj},
         year = 1983,
        month = may,
       volume = {268},
        pages = {368-369},
          doi = {10.1086/160960},
       adsurl = {https://ui.adsabs.harvard.edu/abs/1983ApJ...268..368E}
}

@ARTICLE{jiang2021implicit,
       author = {{Jiang}, Yan-Fei},
        title = "{An Implicit Finite Volume Scheme to Solve the Time-dependent Radiation Transport Equation Based on Discrete Ordinates}",
      journal = {\apjs},
         year = 2021,
        month = apr,
       volume = {253},
       number = {2},
          eid = {49},
        pages = {49},
          doi = {10.3847/1538-4365/abe303},
archivePrefix = {arXiv},
       eprint = {2102.02212},
 primaryClass = {astro-ph.IM},
       adsurl = {https://ui.adsabs.harvard.edu/abs/2021ApJS..253...49J}
}

@ARTICLE{jiang2014explicit,
       author = {{Jiang}, Yan-Fei and {Stone}, James M. and {Davis}, Shane W.},
        title = "{An Algorithm for Radiation Magnetohydrodynamics Based on Solving the Time-dependent Transfer Equation}",
      journal = {\apjs},
         year = 2014,
        month = jul,
       volume = {213},
       number = {1},
          eid = {7},
        pages = {7},
          doi = {10.1088/0067-0049/213/1/7},
archivePrefix = {arXiv},
       eprint = {1403.6126},
 primaryClass = {astro-ph.IM},
       adsurl = {https://ui.adsabs.harvard.edu/abs/2014ApJS..213....7J}
}

@article{davis2014radiation,
  title={Radiation feedback in ULIRGs: are photons movers and shakers?},
  author={Davis, Shane W and Jiang, Yan-Fei and Stone, James M and Murray, Norman},
  journal={The Astrophysical Journal},
  volume={796},
  number={2},
  pages={107},
  year={2014},
  adsurl = {https://ui.adsabs.harvard.edu/abs/2014ApJ...796..107D/abstract},
  publisher={IOP Publishing}
}

@ARTICLE{jiang2019super,
       author = {{Jiang}, Yan-Fei and {Stone}, James M. and {Davis}, Shane W.},
        title = "{Super-Eddington Accretion Disks around Supermassive Black Holes}",
      journal = {\apj},
         year = 2019,
        month = aug,
       volume = {880},
       number = {2},
          eid = {67},
        pages = {67},
          doi = {10.3847/1538-4357/ab29ff},
archivePrefix = {arXiv},
       eprint = {1709.02845},
 primaryClass = {astro-ph.HE},
       adsurl = {https://ui.adsabs.harvard.edu/abs/2019ApJ...880...67J}
}

@article{armengol2021circumbinary,
  title={Circumbinary disk accretion into spinning black hole binaries},
  author={Armengol, Federico G Lopez and Combi, Luciano and Campanelli, Manuela and Noble, Scott C and Krolik, Julian H and Bowen, Dennis B and Avara, Mark J and Mewes, Vassilios and Nakano, Hiroyuki},
  journal={The Astrophysical Journal},
  volume={913},
  number={1},
  pages={16},
  doi={10.3847/1538-4357/abf0af},
  year={2021},
  publisher={IOP Publishing}
}

@ARTICLE{noble2021mass,
       author = {{Noble}, Scott C. and {Krolik}, Julian H. and {Campanelli}, Manuela and {Zlochower}, Yosef and {Mundim}, Bruno C. and {Nakano}, Hiroyuki and {Zilh{\~a}o}, Miguel},
        title = "{Mass-ratio and Magnetic Flux Dependence of Modulated Accretion from Circumbinary Disks}",
      journal = {\apj},
         year = 2021,
        month = dec,
       volume = {922},
       number = {2},
          eid = {175},
        pages = {175},
          doi = {10.3847/1538-4357/ac2229},
archivePrefix = {arXiv},
       eprint = {2103.12100},
 primaryClass = {astro-ph.HE},
       adsurl = {https://ui.adsabs.harvard.edu/abs/2021ApJ...922..175N}
}

@ARTICLE{sudarshan2022cooling,
       author = {{Sudarshan}, Prakruti and {Penzlin}, Anna B.~T. and {Ziampras}, Alexandros and {Kley}, Wilhelm and {Nelson}, Richard P.},
        title = "{How cooling influences circumbinary discs}",
      journal = {\aap},
         year = 2022,
        month = aug,
       volume = {664},
          eid = {A157},
        pages = {A157},
          doi = {10.1051/0004-6361/202243472},
archivePrefix = {arXiv},
       eprint = {2206.07749},
 primaryClass = {astro-ph.EP},
       adsurl = {https://ui.adsabs.harvard.edu/abs/2022A&A...664A.157S}
}

@ARTICLE{wang2023role,
       author = {{Wang}, Hai-Yang and {Bai}, Xue-Ning and {Lai}, Dong},
        title = "{On the Role of Dynamical Cooling in the Dynamics of Circumbinary Disks}",
      journal = {\apj},
         year = 2023,
        month = feb,
       volume = {943},
       number = {2},
          eid = {175},
        pages = {175},
          doi = {10.3847/1538-4357/acac77},
archivePrefix = {arXiv},
       eprint = {2212.04199},
 primaryClass = {astro-ph.HE},
       adsurl = {https://ui.adsabs.harvard.edu/abs/2023ApJ...943..175W}
}

@ARTICLE{munoz2020circumbinary,
       author = {{Mu{\~n}oz}, Diego J. and {Lai}, Dong and {Kratter}, Kaitlin and {Miranda}, Ryan},
        title = "{Circumbinary Accretion from Finite and Infinite Disks}",
      journal = {\apj},
         year = 2020,
        month = feb,
       volume = {889},
       number = {2},
          eid = {114},
        pages = {114},
          doi = {10.3847/1538-4357/ab5d33},
archivePrefix = {arXiv},
       eprint = {1910.04763},
 primaryClass = {astro-ph.HE},
       adsurl = {https://ui.adsabs.harvard.edu/abs/2020ApJ...889..114M}
}

@ARTICLE{heath2020orbital,
       author = {{Heath}, R.~M. and {Nixon}, C.~J.},
        title = "{On the orbital evolution of binaries with circumbinary discs}",
      journal = {\aap},
         year = 2020,
        month = sep,
       volume = {641},
          eid = {A64},
        pages = {A64},
          doi = {10.1051/0004-6361/202038548},
archivePrefix = {arXiv},
       eprint = {2007.11592},
 primaryClass = {astro-ph.HE},
       adsurl = {https://ui.adsabs.harvard.edu/abs/2020A&A...641A..64H}
}

@ARTICLE{thun2017circumbinary,
       author = {{Thun}, Daniel and {Kley}, Wilhelm and {Picogna}, Giovanni},
        title = "{Circumbinary discs: Numerical and physical behaviour}",
      journal = {\aap},
         year = 2017,
        month = aug,
       volume = {604},
          eid = {A102},
        pages = {A102},
          doi = {10.1051/0004-6361/201730666},
archivePrefix = {arXiv},
       eprint = {1704.08130},
 primaryClass = {astro-ph.EP},
       adsurl = {https://ui.adsabs.harvard.edu/abs/2017A&A...604A.102T}
}

@ARTICLE{lai2023circumbinary,
       author = {{Lai}, Dong and {Mu{\~n}oz}, Diego J.},
        title = "{Circumbinary Accretion: From Binary Stars to Massive Binary Black Holes}",
      journal = {\araa},
         year = 2023,
        month = aug,
       volume = {61},
        pages = {517-560},
          doi = {10.1146/annurev-astro-052622-022933},
archivePrefix = {arXiv},
       eprint = {2211.00028},
 primaryClass = {astro-ph.HE},
       adsurl = {https://ui.adsabs.harvard.edu/abs/2023ARA&A..61..517L}
}

@ARTICLE{krauth2023disappearing,
       author = {{Krauth}, Luke Major and {Davelaar}, Jordy and {Haiman}, Zolt{\'a}n and {Westernacher-Schneider}, John Ryan and {Zrake}, Jonathan and {MacFadyen}, Andrew},
        title = "{Disappearing thermal X-ray emission as a tell-tale signature of merging massive black hole binaries}",
      journal = {\mnras},
         year = 2023,
        month = dec,
       volume = {526},
       number = {4},
        pages = {5441-5454},
          doi = {10.1093/mnras/stad3095},
archivePrefix = {arXiv},
       eprint = {2304.02575},
 primaryClass = {astro-ph.HE},
       adsurl = {https://ui.adsabs.harvard.edu/abs/2023MNRAS.526.5441K}
}

@ARTICLE{westernacher2022multiband,
       author = {{Westernacher-Schneider}, John Ryan and {Zrake}, Jonathan and {MacFadyen}, Andrew and {Haiman}, Zolt{\'a}n},
        title = "{Multiband light curves from eccentric accreting supermassive black hole binaries}",
      journal = {\prd},
         year = 2022,
        month = nov,
       volume = {106},
       number = {10},
          eid = {103010},
        pages = {103010},
          doi = {10.1103/PhysRevD.106.103010},
archivePrefix = {arXiv},
       eprint = {2111.06882},
 primaryClass = {astro-ph.HE},
       adsurl = {https://ui.adsabs.harvard.edu/abs/2022PhRvD.106j3010W}
}

@ARTICLE{farris2015binary,
       author = {{Farris}, B.~D. and {Duffell}, P. and {MacFadyen}, A.~I. and {Haiman}, Z.},
        title = "{Binary black hole accretion during inspiral and merger.}",
      journal = {\mnras},
         year = 2015,
        month = feb,
       volume = {447},
        pages = {L80-L84},
          doi = {10.1093/mnrasl/slu184},
archivePrefix = {arXiv},
       eprint = {1409.5124},
 primaryClass = {astro-ph.HE},
       adsurl = {https://ui.adsabs.harvard.edu/abs/2015MNRAS.447L..80F}
}

@ARTICLE{tang2018late,
       author = {{Tang}, Yike and {Haiman}, Zolt{\'a}n and {MacFadyen}, Andrew},
        title = "{The late inspiral of supermassive black hole binaries with circumbinary gas discs in the LISA band}",
      journal = {\mnras},
         year = 2018,
        month = may,
       volume = {476},
       number = {2},
        pages = {2249-2257},
          doi = {10.1093/mnras/sty423},
archivePrefix = {arXiv},
       eprint = {1801.02266},
 primaryClass = {astro-ph.HE},
       adsurl = {https://ui.adsabs.harvard.edu/abs/2018MNRAS.476.2249T}
}

@ARTICLE{franchini2024emission,
       author = {{Franchini}, Alessia and {Bonetti}, Matteo and {Lupi}, Alessandro and {Sesana}, Alberto},
        title = "{Emission signatures from sub-parsec post-Newtonian binaries embedded in circumbinary discs}",
      journal = {\aap},
         year = 2024,
        month = jun,
       volume = {686},
          eid = {A288},
        pages = {A288},
          doi = {10.1051/0004-6361/202449206},
archivePrefix = {arXiv},
       eprint = {2401.10331},
 primaryClass = {astro-ph.HE},
       adsurl = {https://ui.adsabs.harvard.edu/abs/2024A&A...686A.288F}
}

@ARTICLE{tiede2022binaries,
       author = {{Tiede}, Christopher and {Zrake}, Jonathan and {MacFadyen}, Andrew and {Haiman}, Zolt{\'a}n},
        title = "{How Binaries Accrete: Hydrodynamic Simulations with Passive Tracer Particles}",
      journal = {\apj},
         year = 2022,
        month = jun,
       volume = {932},
       number = {1},
          eid = {24},
        pages = {24},
          doi = {10.3847/1538-4357/ac6c2b},
archivePrefix = {arXiv},
       eprint = {2111.04721},
 primaryClass = {astro-ph.GA},
       adsurl = {https://ui.adsabs.harvard.edu/abs/2022ApJ...932...24T}
}

@ARTICLE{zhu2021global,
       author = {{Zhu}, Zhaohuan and {Jiang}, Yan-Fei and {Baehr}, Hans and {Youdin}, Andrew N. and {Armitage}, Philip J. and {Martin}, Rebecca G.},
        title = "{Global 3D radiation hydrodynamic simulations of proto-Jupiter's convective envelope}",
      journal = {\mnras},
         year = 2021,
        month = nov,
       volume = {508},
       number = {1},
        pages = {453-474},
          doi = {10.1093/mnras/stab2517},
archivePrefix = {arXiv},
       eprint = {2106.12003},
 primaryClass = {astro-ph.EP},
       adsurl = {https://ui.adsabs.harvard.edu/abs/2021MNRAS.508..453Z}
}

@BOOK{rybicki1991radiative,
       author = {{Rybicki}, George B. and {Lightman}, Alan P.},
        title = "{Radiative processes in astrophysics}",
         year = 1979,
       adsurl = {https://ui.adsabs.harvard.edu/abs/1979rpa..book.....R}
}

@article{agazie2023nanograv_gwbackground,
  title={The NANOGrav 15 yr data set: Evidence for a gravitational-wave background},
  author={Agazie, Gabriella and Anumarlapudi, Akash and Archibald, Anne M and Arzoumanian, Zaven and Baker, Paul T and B{\'e}csy, Bence and Blecha, Laura and Brazier, Adam and Brook, Paul R and Burke-Spolaor, Sarah and others},
  journal={The Astrophysical Journal Letters},
  volume={951},
  number={1},
  pages={L8},
  doi = {10.3847/2041-8213/acdac6},
  year={2023},
  publisher={IOP Publishing}
}

@article{antoniadis2024second,
  title={The second data release from the European Pulsar Timing Array-IV. Implications for massive black holes, dark matter, and the early Universe},
  author={Antoniadis, J and Arumugam, P and Arumugam, S and Babak, S and Bagchi, M and Nielsen, A-S Bak and Bassa, CG and Bathula, A and Berthereau, A and Bonetti, M and others},
  journal={Astronomy \& Astrophysics},
  volume={685},
  pages={A94},
  doi={10.1051/0004-6361/202347433},
  year={2024},
  publisher={EDP Sciences}
}

@ARTICLE{reardon2023search,
       author = {{Reardon}, Daniel J. and {Zic}, Andrew and {Shannon}, Ryan M. and {Hobbs}, George B. and {Bailes}, Matthew and {Di Marco}, Valentina and {Kapur}, Agastya and {Rogers}, Axl F. and {Thrane}, Eric and {Askew}, Jacob and {Bhat}, N.~D. Ramesh and {Cameron}, Andrew and {Cury{\l}o}, Ma{\l}gorzata and {Coles}, William A. and {Dai}, Shi and {Goncharov}, Boris and {Kerr}, Matthew and {Kulkarni}, Atharva and {Levin}, Yuri and {Lower}, Marcus E. and {Manchester}, Richard N. and {Mandow}, Rami and {Miles}, Matthew T. and {Nathan}, Rowina S. and {Os{\l}owski}, Stefan and {Russell}, Christopher J. and {Spiewak}, Ren{\'e}e and {Zhang}, Songbo and {Zhu}, Xing-Jiang},
        title = "{Search for an Isotropic Gravitational-wave Background with the Parkes Pulsar Timing Array}",
      journal = {\apjl},
         year = 2023,
        month = jul,
       volume = {951},
       number = {1},
          eid = {L6},
        pages = {L6},
          doi = {10.3847/2041-8213/acdd02},
archivePrefix = {arXiv},
       eprint = {2306.16215},
 primaryClass = {astro-ph.HE},
       adsurl = {https://ui.adsabs.harvard.edu/abs/2023ApJ...951L...6R}
}

@ARTICLE{xu2023searching,
       author = {{Xu}, Heng and {Chen}, Siyuan and {Guo}, Yanjun and {Jiang}, Jinchen and {Wang}, Bojun and {Xu}, Jiangwei and {Xue}, Zihan and {Caballero}, R. Nicolas and {Yuan}, Jianping and {Xu}, Yonghua and {Wang}, Jingbo and {Hao}, Longfei and {Luo}, Jingtao and {Lee}, Kejia and {Han}, Jinlin and {Jiang}, Peng and {Shen}, Zhiqiang and {Wang}, Min and {Wang}, Na and {Xu}, Renxin and {Wu}, Xiangping and {Manchester}, Richard and {Qian}, Lei and {Guan}, Xin and {Huang}, Menglin and {Sun}, Chun and {Zhu}, Yan},
        title = "{Searching for the Nano-Hertz Stochastic Gravitational Wave Background with the Chinese Pulsar Timing Array Data Release I}",
      journal = {Research in Astronomy and Astrophysics},
         year = 2023,
        month = jul,
       volume = {23},
       number = {7},
          eid = {075024},
        pages = {075024},
          doi = {10.1088/1674-4527/acdfa5},
archivePrefix = {arXiv},
       eprint = {2306.16216},
 primaryClass = {astro-ph.HE},
       adsurl = {https://ui.adsabs.harvard.edu/abs/2023RAA....23g5024X}
}

@article{amaro2023astrophysics,
  title={Astrophysics with the laser interferometer space antenna},
  author={Amaro-Seoane, Pau and Andrews, Jeff and Arca Sedda, Manuel and Askar, Abbas and Baghi, Quentin and Balasov, Razvan and Bartos, Imre and Bavera, Simone S and Bellovary, Jillian and Berry, Christopher PL and others},
  journal={Living Reviews in Relativity},
  volume={26},
  number={1},
  pages={2},
  doi={10.1007/s41114-022-00041-y},
  year={2023},
  publisher={Springer}
}

@ARTICLE{turpin2024orbital,
       author = {{Turpin}, George A. and {Nelson}, Richard P.},
        title = "{Orbital evolution of close binary systems: comparing viscous and wind-driven circumbinary disc models}",
      journal = {\mnras},
         year = 2024,
        month = mar,
       volume = {528},
       number = {4},
        pages = {7256-7273},
          doi = {10.1093/mnras/stae109},
archivePrefix = {arXiv},
       eprint = {2401.10808},
 primaryClass = {astro-ph.EP},
       adsurl = {https://ui.adsabs.harvard.edu/abs/2024MNRAS.528.7256T}
}

@ARTICLE{tiede2024eccentric,
       author = {{Tiede}, Christopher and {D'Orazio}, Daniel J.},
        title = "{Eccentric binaries in retrograde discs}",
      journal = {\mnras},
         year = 2024,
        month = jan,
       volume = {527},
       number = {3},
        pages = {6021-6037},
          doi = {10.1093/mnras/stad3551},
archivePrefix = {arXiv},
       eprint = {2307.03775},
 primaryClass = {astro-ph.GA},
       adsurl = {https://ui.adsabs.harvard.edu/abs/2024MNRAS.527.6021T}
}

@ARTICLE{dittmann2023decoupling,
       author = {{Dittmann}, Alexander J. and {Ryan}, Geoffrey and {Miller}, M. Coleman},
        title = "{The Decoupling of Binaries from Their Circumbinary Disks}",
      journal = {\apjl},
         year = 2023,
        month = jun,
       volume = {949},
       number = {2},
          eid = {L30},
        pages = {L30},
          doi = {10.3847/2041-8213/acd183},
archivePrefix = {arXiv},
       eprint = {2303.16204},
 primaryClass = {astro-ph.HE},
       adsurl = {https://ui.adsabs.harvard.edu/abs/2023ApJ...949L..30D}
}

@ARTICLE{siwek2023preferential,
       author = {{Siwek}, Magdalena and {Weinberger}, Rainer and {Mu{\~n}oz}, Diego J. and {Hernquist}, Lars},
        title = "{Preferential accretion and circumbinary disc precession in eccentric binary systems}",
      journal = {\mnras},
         year = 2023,
        month = feb,
       volume = {518},
       number = {4},
        pages = {5059-5071},
          doi = {10.1093/mnras/stac3263},
archivePrefix = {arXiv},
       eprint = {2203.02514},
 primaryClass = {astro-ph.HE},
       adsurl = {https://ui.adsabs.harvard.edu/abs/2023MNRAS.518.5059S}
}

@ARTICLE{franchini2023importance,
       author = {{Franchini}, Alessia and {Lupi}, Alessandro and {Sesana}, Alberto and {Haiman}, Zoltan},
        title = "{The importance of live binary evolution in numerical simulations of binaries embedded in circumbinary discs}",
      journal = {\mnras},
         year = 2023,
        month = jun,
       volume = {522},
       number = {1},
        pages = {1569-1574},
          doi = {10.1093/mnras/stad1070},
archivePrefix = {arXiv},
       eprint = {2304.03790},
 primaryClass = {astro-ph.HE},
       adsurl = {https://ui.adsabs.harvard.edu/abs/2023MNRAS.522.1569F}
}

@ARTICLE{smallwood2022accretion,
       author = {{Smallwood}, Jeremy L. and {Lubow}, Stephen H. and {Martin}, Rebecca G.},
        title = "{Accretion on to a binary from a polar circumbinary disc}",
      journal = {\mnras},
         year = 2022,
        month = jul,
       volume = {514},
       number = {1},
        pages = {1249-1257},
          doi = {10.1093/mnras/stac1416},
archivePrefix = {arXiv},
       eprint = {2205.09183},
 primaryClass = {astro-ph.EP},
       adsurl = {https://ui.adsabs.harvard.edu/abs/2022MNRAS.514.1249S}
}

@ARTICLE{zrake2021equilibrium,
       author = {{Zrake}, Jonathan and {Tiede}, Christopher and {MacFadyen}, Andrew and {Haiman}, Zolt{\'a}n},
        title = "{Equilibrium Eccentricity of Accreting Binaries}",
      journal = {\apjl},
         year = 2021,
        month = mar,
       volume = {909},
       number = {1},
          eid = {L13},
        pages = {L13},
          doi = {10.3847/2041-8213/abdd1c},
archivePrefix = {arXiv},
       eprint = {2010.09707},
 primaryClass = {astro-ph.HE},
       adsurl = {https://ui.adsabs.harvard.edu/abs/2021ApJ...909L..13Z}
}

@ARTICLE{d2021orbital,
       author = {{D'Orazio}, Daniel J. and {Duffell}, Paul C.},
        title = "{Orbital Evolution of Equal-mass Eccentric Binaries due to a Gas Disk: Eccentric Inspirals and Circular Outspirals}",
      journal = {\apjl},
         year = 2021,
        month = jun,
       volume = {914},
       number = {1},
          eid = {L21},
        pages = {L21},
          doi = {10.3847/2041-8213/ac0621},
archivePrefix = {arXiv},
       eprint = {2103.09251},
 primaryClass = {astro-ph.HE},
       adsurl = {https://ui.adsabs.harvard.edu/abs/2021ApJ...914L..21D}
}

@ARTICLE{franchini2021circumbinary,
       author = {{Franchini}, Alessia and {Sesana}, Alberto and {Dotti}, Massimo},
        title = "{Circumbinary disc self-gravity governing supermassive black hole binary mergers}",
      journal = {\mnras},
         year = 2021,
        month = oct,
       volume = {507},
       number = {1},
        pages = {1458-1467},
          doi = {10.1093/mnras/stab2234},
archivePrefix = {arXiv},
       eprint = {2106.13253},
 primaryClass = {astro-ph.HE},
       adsurl = {https://ui.adsabs.harvard.edu/abs/2021MNRAS.507.1458F}
}

@article{duffell2020circumbinary,
  title={Circumbinary disks: accretion and torque as a function of mass ratio and disk viscosity},
  author={Duffell, Paul C and D’Orazio, Daniel and Derdzinski, Andrea and Haiman, Zoltan and MacFadyen, Andrew and Rosen, Anna L and Zrake, Jonathan},
  journal={The Astrophysical Journal},
  volume={901},
  number={1},
  pages={25},
  doi={10.3847/1538-4357/abab95},
  year={2020},
  publisher={IOP Publishing}
}

@ARTICLE{ragusa2020evolution,
       author = {{Ragusa}, Enrico and {Alexander}, Richard and {Calcino}, Josh and {Hirsh}, Kieran and {Price}, Daniel J.},
        title = "{The evolution of large cavities and disc eccentricity in circumbinary discs}",
      journal = {\mnras},
         year = 2020,
        month = dec,
       volume = {499},
       number = {3},
        pages = {3362-3380},
          doi = {10.1093/mnras/staa2954},
archivePrefix = {arXiv},
       eprint = {2009.10738},
 primaryClass = {astro-ph.EP},
       adsurl = {https://ui.adsabs.harvard.edu/abs/2020MNRAS.499.3362R}
}

@ARTICLE{hirsh2020cavity,
       author = {{Hirsh}, Kieran and {Price}, Daniel J. and {Gonzalez}, Jean-Fran{\c{c}}ois and {Ubeira-Gabellini}, M. Giulia and {Ragusa}, Enrico},
        title = "{On the cavity size in circumbinary discs}",
      journal = {\mnras},
         year = 2020,
        month = oct,
       volume = {498},
       number = {2},
        pages = {2936-2947},
          doi = {10.1093/mnras/staa2536},
archivePrefix = {arXiv},
       eprint = {2008.08008},
 primaryClass = {astro-ph.EP},
       adsurl = {https://ui.adsabs.harvard.edu/abs/2020MNRAS.498.2936H}
}

@ARTICLE{bowen2017relativistic,
       author = {{Bowen}, Dennis B. and {Campanelli}, Manuela and {Krolik}, Julian H. and {Mewes}, Vassilios and {Noble}, Scott C.},
        title = "{Relativistic Dynamics and Mass Exchange in Binary Black Hole Mini-disks}",
      journal = {\apj},
         year = 2017,
        month = mar,
       volume = {838},
       number = {1},
          eid = {42},
        pages = {42},
          doi = {10.3847/1538-4357/aa63f3},
archivePrefix = {arXiv},
       eprint = {1612.02373},
 primaryClass = {astro-ph.HE},
       adsurl = {https://ui.adsabs.harvard.edu/abs/2017ApJ...838...42B}
}

@ARTICLE{ryan2017minidisks,
       author = {{Ryan}, Geoffrey and {MacFadyen}, Andrew},
        title = "{Minidisks in Binary Black Hole Accretion}",
      journal = {\apj},
         year = 2017,
        month = feb,
       volume = {835},
       number = {2},
          eid = {199},
        pages = {199},
          doi = {10.3847/1538-4357/835/2/199},
archivePrefix = {arXiv},
       eprint = {1611.00341},
 primaryClass = {astro-ph.HE},
       adsurl = {https://ui.adsabs.harvard.edu/abs/2017ApJ...835..199R}
}

@ARTICLE{d2016transition,
       author = {{D'Orazio}, Daniel J. and {Haiman}, Zolt{\'a}n and {Duffell}, Paul and {MacFadyen}, Andrew and {Farris}, Brian},
        title = "{A transition in circumbinary accretion discs at a binary mass ratio of 1:25}",
      journal = {\mnras},
         year = 2016,
        month = jul,
       volume = {459},
       number = {3},
        pages = {2379-2393},
          doi = {10.1093/mnras/stw792},
archivePrefix = {arXiv},
       eprint = {1512.05788},
 primaryClass = {astro-ph.HE},
       adsurl = {https://ui.adsabs.harvard.edu/abs/2016MNRAS.459.2379D}
}

@ARTICLE{munoz2016pulsed,
       author = {{Mu{\~n}oz}, Diego J. and {Lai}, Dong},
        title = "{Pulsed Accretion onto Eccentric and Circular Binaries}",
      journal = {\apj},
         year = 2016,
        month = aug,
       volume = {827},
       number = {1},
          eid = {43},
        pages = {43},
          doi = {10.3847/0004-637X/827/1/43},
archivePrefix = {arXiv},
       eprint = {1604.00004},
 primaryClass = {astro-ph.EP},
       adsurl = {https://ui.adsabs.harvard.edu/abs/2016ApJ...827...43M}
}

@ARTICLE{ragusa2016suppression,
       author = {{Ragusa}, Enrico and {Lodato}, Giuseppe and {Price}, Daniel J.},
        title = "{Suppression of the accretion rate in thin discs around binary black holes}",
      journal = {\mnras},
         year = 2016,
        month = aug,
       volume = {460},
       number = {2},
        pages = {1243-1253},
          doi = {10.1093/mnras/stw1081},
archivePrefix = {arXiv},
       eprint = {1605.01730},
 primaryClass = {astro-ph.HE},
       adsurl = {https://ui.adsabs.harvard.edu/abs/2016MNRAS.460.1243R}
}

@ARTICLE{farris2015characteristic,
       author = {{Farris}, B.~D. and {Duffell}, P. and {MacFadyen}, A.~I. and {Haiman}, Z.},
        title = "{Characteristic signatures in the thermal emission from accreting binary black holes.}",
      journal = {\mnras},
         year = 2015,
        month = jan,
       volume = {446},
        pages = {L36-L40},
          doi = {10.1093/mnrasl/slu160},
archivePrefix = {arXiv},
       eprint = {1406.0007},
 primaryClass = {astro-ph.HE},
       adsurl = {https://ui.adsabs.harvard.edu/abs/2015MNRAS.446L..36F}
}

@ARTICLE{young2015binary,
       author = {{Young}, M.~D. and {Clarke}, C.~J.},
        title = "{Binary accretion rates: dependence on temperature and mass ratio}",
      journal = {\mnras},
         year = 2015,
        month = sep,
       volume = {452},
       number = {3},
        pages = {3085-3091},
          doi = {10.1093/mnras/stv1512},
archivePrefix = {arXiv},
       eprint = {1507.01850},
 primaryClass = {astro-ph.SR},
       adsurl = {https://ui.adsabs.harvard.edu/abs/2015MNRAS.452.3085Y}
}

@ARTICLE{nixon2015resonances,
       author = {{Nixon}, Chris and {Lubow}, Stephen H.},
        title = "{Resonances in retrograde circumbinary discs}",
      journal = {\mnras},
         year = 2015,
        month = apr,
       volume = {448},
       number = {4},
        pages = {3472-3483},
          doi = {10.1093/mnras/stv166},
archivePrefix = {arXiv},
       eprint = {1501.07277},
 primaryClass = {astro-ph.HE},
       adsurl = {https://ui.adsabs.harvard.edu/abs/2015MNRAS.448.3472N}
}

@ARTICLE{roedig2014migration,
       author = {{Roedig}, Constanze and {Sesana}, Alberto},
        title = "{Migration of massive black hole binaries in self-gravitating discs: retrograde versus prograde}",
      journal = {\mnras},
         year = 2014,
        month = apr,
       volume = {439},
       number = {4},
        pages = {3476-3489},
          doi = {10.1093/mnras/stu194},
archivePrefix = {arXiv},
       eprint = {1307.6283},
 primaryClass = {astro-ph.HE},
       adsurl = {https://ui.adsabs.harvard.edu/abs/2014MNRAS.439.3476R}
}

@ARTICLE{roedig2012evolution,
       author = {{Roedig}, C. and {Sesana}, A. and {Dotti}, M. and {Cuadra}, J. and {Amaro-Seoane}, P. and {Haardt}, F.},
        title = "{Evolution of binary black holes in self gravitating discs. Dissecting the torques}",
      journal = {\aap},
         year = 2012,
        month = sep,
       volume = {545},
          eid = {A127},
        pages = {A127},
          doi = {10.1051/0004-6361/201219986},
archivePrefix = {arXiv},
       eprint = {1202.6063},
 primaryClass = {astro-ph.CO},
       adsurl = {https://ui.adsabs.harvard.edu/abs/2012A&A...545A.127R}
}

@ARTICLE{roedig2011limiting,
       author = {{Roedig}, C. and {Dotti}, M. and {Sesana}, A. and {Cuadra}, J. and {Colpi}, M.},
        title = "{Limiting eccentricity of subparsec massive black hole binaries surrounded by self-gravitating gas discs}",
      journal = {\mnras},
         year = 2011,
        month = aug,
       volume = {415},
       number = {4},
        pages = {3033-3041},
          doi = {10.1111/j.1365-2966.2011.18927.x},
archivePrefix = {arXiv},
       eprint = {1104.3868},
 primaryClass = {astro-ph.CO},
       adsurl = {https://ui.adsabs.harvard.edu/abs/2011MNRAS.415.3033R}
}

@ARTICLE{gold2014accretion,
       author = {{Gold}, Roman and {Paschalidis}, Vasileios and {Etienne}, Zachariah B. and {Shapiro}, Stuart L. and {Pfeiffer}, Harald P.},
        title = "{Accretion disks around binary black holes of unequal mass: General relativistic magnetohydrodynamic simulations near decoupling}",
      journal = {\prd},
         year = 2014,
        month = mar,
       volume = {89},
       number = {6},
          eid = {064060},
        pages = {064060},
          doi = {10.1103/PhysRevD.89.064060},
archivePrefix = {arXiv},
       eprint = {1312.0600},
 primaryClass = {astro-ph.HE},
       adsurl = {https://ui.adsabs.harvard.edu/abs/2014PhRvD..89f4060G}
}

@ARTICLE{gold2014accretion_two,
       author = {{Gold}, Roman and {Paschalidis}, Vasileios and {Ruiz}, Milton and {Shapiro}, Stuart L. and {Etienne}, Zachariah B. and {Pfeiffer}, Harald P.},
        title = "{Accretion disks around binary black holes of unequal mass: General relativistic MHD simulations of postdecoupling and merger}",
      journal = {\prd},
         year = 2014,
        month = nov,
       volume = {90},
       number = {10},
          eid = {104030},
        pages = {104030},
          doi = {10.1103/PhysRevD.90.104030},
archivePrefix = {arXiv},
       eprint = {1410.1543},
 primaryClass = {astro-ph.GA},
       adsurl = {https://ui.adsabs.harvard.edu/abs/2014PhRvD..90j4030G}
}

@ARTICLE{farris2012binary,
       author = {{Farris}, Brian D. and {Gold}, Roman and {Paschalidis}, Vasileios and {Etienne}, Zachariah B. and {Shapiro}, Stuart L.},
        title = "{Binary Black-Hole Mergers in Magnetized Disks: Simulations in Full General Relativity}",
      journal = {\prl},
         year = 2012,
        month = nov,
       volume = {109},
       number = {22},
          eid = {221102},
        pages = {221102},
          doi = {10.1103/PhysRevLett.109.221102},
archivePrefix = {arXiv},
       eprint = {1207.3354},
 primaryClass = {astro-ph.HE},
       adsurl = {https://ui.adsabs.harvard.edu/abs/2012PhRvL.109v1102F}
}

@ARTICLE{bowen2018quasi,
       author = {{Bowen}, Dennis B. and {Mewes}, Vassilios and {Campanelli}, Manuela and {Noble}, Scott C. and {Krolik}, Julian H. and {Zilh{\~a}o}, Miguel},
        title = "{Quasi-periodic Behavior of Mini-disks in Binary Black Holes Approaching Merger}",
      journal = {\apjl},
         year = 2018,
        month = jan,
       volume = {853},
       number = {1},
          eid = {L17},
        pages = {L17},
          doi = {10.3847/2041-8213/aaa756},
archivePrefix = {arXiv},
       eprint = {1712.05451},
 primaryClass = {astro-ph.HE},
       adsurl = {https://ui.adsabs.harvard.edu/abs/2018ApJ...853L..17B}
}

@ARTICLE{bowen2019quasi,
       author = {{Bowen}, Dennis B. and {Mewes}, Vassilios and {Noble}, Scott C. and {Avara}, Mark and {Campanelli}, Manuela and {Krolik}, Julian H.},
        title = "{Quasi-periodicity of Supermassive Binary Black Hole Accretion Approaching Merger}",
      journal = {\apj},
         year = 2019,
        month = jul,
       volume = {879},
       number = {2},
          eid = {76},
        pages = {76},
          doi = {10.3847/1538-4357/ab2453},
archivePrefix = {arXiv},
       eprint = {1904.12048},
 primaryClass = {astro-ph.HE},
       adsurl = {https://ui.adsabs.harvard.edu/abs/2019ApJ...879...76B}
}

@ARTICLE{paschalidis2021minidisk,
       author = {{Paschalidis}, Vasileios and {Bright}, Jane and {Ruiz}, Milton and {Gold}, Roman},
        title = "{Minidisk Dynamics in Accreting, Spinning Black Hole Binaries: Simulations in Full General Relativity}",
      journal = {\apjl},
         year = 2021,
        month = apr,
       volume = {910},
       number = {2},
          eid = {L26},
        pages = {L26},
          doi = {10.3847/2041-8213/abee21},
archivePrefix = {arXiv},
       eprint = {2102.06712},
 primaryClass = {astro-ph.HE},
       adsurl = {https://ui.adsabs.harvard.edu/abs/2021ApJ...910L..26P}
}

@ARTICLE{cattorini2021fully,
       author = {{Cattorini}, Federico and {Giacomazzo}, Bruno and {Haardt}, Francesco and {Colpi}, Monica},
        title = "{Fully general relativistic magnetohydrodynamic simulations of accretion flows onto spinning massive black hole binary mergers}",
      journal = {\prd},
         year = 2021,
        month = may,
       volume = {103},
       number = {10},
          eid = {103022},
        pages = {103022},
          doi = {10.1103/PhysRevD.103.103022},
archivePrefix = {arXiv},
       eprint = {2102.13166},
 primaryClass = {astro-ph.HE},
       adsurl = {https://ui.adsabs.harvard.edu/abs/2021PhRvD.103j3022C}
}

@article{combi2022minidisk,
  title={Minidisk Accretion onto Spinning Black Hole Binaries: Quasi-periodicities and Outflows},
  author={Combi, Luciano and Armengol, Federico G Lopez and Campanelli, Manuela and Noble, Scott C and Avara, Mark and Krolik, Julian H and Bowen, Dennis},
  journal={The Astrophysical Journal},
  volume={928},
  number={2},
  pages={187},
  doi={10.3847/1538-4357/ac532a},
  year={2022},
  publisher={IOP Publishing}
}

@article{avara2023accretion,
  title={Accretion onto a supermassive black hole binary before merger},
  author={Avara, Mark J and Krolik, Julian H and Campanelli, Manuela and Noble, Scott C and Bowen, Dennis and Ryu, Taeho},
  journal={arXiv preprint arXiv:2305.18538},
  year={2023}
}

@ARTICLE{bright2023minidisc,
       author = {{Bright}, Jane C. and {Paschalidis}, Vasileios},
        title = "{Minidisc influence on flow variability in accreting spinning black hole binaries: simulations in full general relativity}",
      journal = {\mnras},
         year = 2023,
        month = mar,
       volume = {520},
       number = {1},
        pages = {392-401},
          doi = {10.1093/mnras/stad091},
archivePrefix = {arXiv},
       eprint = {2210.15686},
 primaryClass = {astro-ph.HE},
       adsurl = {https://ui.adsabs.harvard.edu/abs/2023MNRAS.520..392B}
}

@ARTICLE{ruiz2023general,
       author = {{Ruiz}, Milton and {Tsokaros}, Antonios and {Shapiro}, Stuart L.},
        title = "{General relativistic magnetohydrodynamic simulations of accretion disks around tilted binary black holes of unequal mass}",
      journal = {\prd},
         year = 2023,
        month = dec,
       volume = {108},
       number = {12},
          eid = {124043},
        pages = {124043},
          doi = {10.1103/PhysRevD.108.124043},
archivePrefix = {arXiv},
       eprint = {2302.09083},
 primaryClass = {astro-ph.HE},
       adsurl = {https://ui.adsabs.harvard.edu/abs/2023PhRvD.108l4043R}
}

@ARTICLE{giacomazzo2012general,
       author = {{Giacomazzo}, Bruno and {Baker}, John G. and {Miller}, M. Coleman and {Reynolds}, Christopher S. and {van Meter}, James R.},
        title = "{General Relativistic Simulations of Magnetized Plasmas around Merging Supermassive Black Holes}",
      journal = {\apjl},
         year = 2012,
        month = jun,
       volume = {752},
       number = {1},
          eid = {L15},
        pages = {L15},
          doi = {10.1088/2041-8205/752/1/L15},
archivePrefix = {arXiv},
       eprint = {1203.6108},
 primaryClass = {astro-ph.HE},
       adsurl = {https://ui.adsabs.harvard.edu/abs/2012ApJ...752L..15G}
}

@ARTICLE{kelly2017prompt,
       author = {{Kelly}, Bernard J. and {Baker}, John G. and {Etienne}, Zachariah B. and {Giacomazzo}, Bruno and {Schnittman}, Jeremy},
        title = "{Prompt electromagnetic transients from binary black hole mergers}",
      journal = {\prd},
         year = 2017,
        month = dec,
       volume = {96},
       number = {12},
          eid = {123003},
        pages = {123003},
          doi = {10.1103/PhysRevD.96.123003},
archivePrefix = {arXiv},
       eprint = {1710.02132},
 primaryClass = {astro-ph.HE},
       adsurl = {https://ui.adsabs.harvard.edu/abs/2017PhRvD..96l3003K}
}

@Article{Hunter:2007,
  Author    = {Hunter, J. D.},
  Title     = {Matplotlib: A 2D graphics environment},
  Journal   = {Computing in Science \& Engineering},
  Volume    = {9},
  Number    = {3},
  Pages     = {90--95},
  publisher = {IEEE COMPUTER SOC},
  doi       = {10.1109/MCSE.2007.55},
  year      = 2007
}

@article{dalcin2005mpi,
  title={MPI for Python},
  author={Dalc{\'\i}n, Lisandro and Paz, Rodrigo and Storti, Mario},
  journal={Journal of Parallel and Distributed Computing},
  volume={65},
  number={9},
  pages={1108--1115},
  doi={10.1016/j.jpdc.2007.09.005},
  year={2005},
  publisher={Elsevier}
}

@Article{         harris2020array,
 title         = {Array programming with {NumPy}},
 author        = {Charles R. Harris and K. Jarrod Millman and St{\'{e}}fan J.
                 van der Walt and Ralf Gommers and Pauli Virtanen and David
                 Cournapeau and Eric Wieser and Julian Taylor and Sebastian
                 Berg and Nathaniel J. Smith and Robert Kern and Matti Picus
                 and Stephan Hoyer and Marten H. van Kerkwijk and Matthew
                 Brett and Allan Haldane and Jaime Fern{\'{a}}ndez del
                 R{\'{i}}o and Mark Wiebe and Pearu Peterson and Pierre
                 G{\'{e}}rard-Marchant and Kevin Sheppard and Tyler Reddy and
                 Warren Weckesser and Hameer Abbasi and Christoph Gohlke and
                 Travis E. Oliphant},
 year          = {2020},
 month         = sep,
 journal       = {Nature},
 volume        = {585},
 number        = {7825},
 pages         = {357--362},
 doi           = {10.1038/s41586-020-2649-2},
 publisher     = {Springer Science and Business Media {LLC}},
 url           = {https://doi.org/10.1038/s41586-020-2649-2}
}

@ARTICLE{hawley2013testing,
       author = {{Hawley}, John F. and {Richers}, Sherwood A. and {Guan}, Xiaoyue and {Krolik}, Julian H.},
        title = "{Testing Convergence for Global Accretion Disks}",
      journal = {\apj},
         year = 2013,
        month = aug,
       volume = {772},
       number = {2},
          eid = {102},
        pages = {102},
          doi = {10.1088/0004-637X/772/2/102},
archivePrefix = {arXiv},
       eprint = {1306.0243},
 primaryClass = {astro-ph.IM},
       adsurl = {https://ui.adsabs.harvard.edu/abs/2013ApJ...772..102H}
}

@ARTICLE{sorathia2012global,
       author = {{Sorathia}, Kareem A. and {Reynolds}, Christopher S. and {Stone}, James M. and {Beckwith}, Kris},
        title = "{Global Simulations of Accretion Disks. I. Convergence and Comparisons with Local Models}",
      journal = {\apj},
         year = 2012,
        month = apr,
       volume = {749},
       number = {2},
          eid = {189},
        pages = {189},
          doi = {10.1088/0004-637X/749/2/189},
archivePrefix = {arXiv},
       eprint = {1106.4019},
 primaryClass = {astro-ph.HE},
       adsurl = {https://ui.adsabs.harvard.edu/abs/2012ApJ...749..189S}
}

@article{agazie2023nanograv,
  title={The NANOGrav 15 yr data set: Bayesian limits on gravitational waves from individual supermassive black hole binaries},
  author={Agazie, Gabriella and Anumarlapudi, Akash and Archibald, Anne M and Arzoumanian, Zaven and Baker, Paul T and B{\'e}csy, Bence and Blecha, Laura and Brazier, Adam and Brook, Paul R and Burke-Spolaor, Sarah and others},
  journal={The Astrophysical Journal Letters},
  volume={951},
  number={2},
  pages={L50},
  year={2023},
  publisher={IOP Publishing},
  doi={10.3847/2041-8213/ace18a}
}

@article{agazie2024nanograv,
  title={The NANOGrav 12.5 yr Data Set: A Computationally Efficient Eccentric Binary Search Pipeline and Constraints on an Eccentric Supermassive Binary Candidate in 3C 66B},
  author={Agazie, Gabriella and Arzoumanian, Zaven and Baker, Paul T and B{\'e}csy, Bence and Blecha, Laura and Blumer, Harsha and Brazier, Adam and Brook, Paul R and Burke-Spolaor, Sarah and Casey-Clyde, J Andrew and others},
  journal={The Astrophysical Journal},
  volume={963},
  number={2},
  pages={144},
  doi={10.3847/1538-4357/ad1f61},
  year={2024},
  publisher={IOP Publishing}
}

@ARTICLE{most2024magnetically,
       author = {{Most}, Elias R. and {Wang}, Hai-Yang},
        title = "{Magnetically Arrested Circumbinary Accretion Flows}",
      journal = {\apjl},
         year = 2024,
        month = sep,
       volume = {973},
       number = {1},
          eid = {L19},
        pages = {L19},
          doi = {10.3847/2041-8213/ad7713},
archivePrefix = {arXiv},
       eprint = {2408.00757},
 primaryClass = {astro-ph.HE},
       adsurl = {https://ui.adsabs.harvard.edu/abs/2024ApJ...973L..19M}
}

@ARTICLE{mills2024spectral,
       author = {{Mills}, Brianna S. and {Davis}, Shane W. and {Jiang}, Yan-Fei and {Middleton}, Matthew J.},
        title = "{Spectral Calculations of 3D Radiation Magnetohydrodynamic Simulations of Super-Eddington Accretion onto a Stellar-mass Black Hole}",
      journal = {\apj},
         year = 2024,
        month = oct,
       volume = {974},
       number = {2},
          eid = {166},
        pages = {166},
          doi = {10.3847/1538-4357/ad6b21},
archivePrefix = {arXiv},
       eprint = {2304.07977},
 primaryClass = {astro-ph.HE},
       adsurl = {https://ui.adsabs.harvard.edu/abs/2024ApJ...974..166M}
}

@article{shvartzvald2023ultrasat,
  title={ULTRASAT: A wide-field time-domain UV space telescope, arXiv e-prints},
  author={Shvartzvald, Y and Waxman, E and Gal-Yam, A and Ofek, EO and Ben-Ami, S and Berge, D and others},
  journal={arXiv preprint arXiv:2304.14482},
  year={2023}
}

@ARTICLE{ivezic2019lsst,
       author = {{Ivezi{\'c}}, {\v{Z}}eljko and {Kahn}, Steven M. and {Tyson}, J. Anthony and {Abel}, Bob and {Acosta}, Emily and {Allsman}, Robyn and {Alonso}, David and {AlSayyad}, Yusra and {Anderson}, Scott F. and {Andrew}, John and {Angel}, James Roger P. and {Angeli}, George Z. and {Ansari}, Reza and {Antilogus}, Pierre and {Araujo}, Constanza and {Armstrong}, Robert and {Arndt}, Kirk T. and {Astier}, Pierre and {Aubourg}, {\'E}ric and {Auza}, Nicole and {Axelrod}, Tim S. and {Bard}, Deborah J. and {Barr}, Jeff D. and {Barrau}, Aurelian and {Bartlett}, James G. and {Bauer}, Amanda E. and {Bauman}, Brian J. and {Baumont}, Sylvain and {Bechtol}, Ellen and {Bechtol}, Keith and {Becker}, Andrew C. and {Becla}, Jacek and {Beldica}, Cristina and {Bellavia}, Steve and {Bianco}, Federica B. and {Biswas}, Rahul and {Blanc}, Guillaume and {Blazek}, Jonathan and {Blandford}, Roger D. and {Bloom}, Josh S. and {Bogart}, Joanne and {Bond}, Tim W. and {Booth}, Michael T. and {Borgland}, Anders W. and {Borne}, Kirk and {Bosch}, James F. and {Boutigny}, Dominique and {Brackett}, Craig A. and {Bradshaw}, Andrew and {Brandt}, William Nielsen and {Brown}, Michael E. and {Bullock}, James S. and {Burchat}, Patricia and {Burke}, David L. and {Cagnoli}, Gianpietro and {Calabrese}, Daniel and {Callahan}, Shawn and {Callen}, Alice L. and {Carlin}, Jeffrey L. and {Carlson}, Erin L. and {Chandrasekharan}, Srinivasan and {Charles-Emerson}, Glenaver and {Chesley}, Steve and {Cheu}, Elliott C. and {Chiang}, Hsin-Fang and {Chiang}, James and {Chirino}, Carol and {Chow}, Derek and {Ciardi}, David R. and {Claver}, Charles F. and {Cohen-Tanugi}, Johann and {Cockrum}, Joseph J. and {Coles}, Rebecca and {Connolly}, Andrew J. and {Cook}, Kem H. and {Cooray}, Asantha and {Covey}, Kevin R. and {Cribbs}, Chris and {Cui}, Wei and {Cutri}, Roc and {Daly}, Philip N. and {Daniel}, Scott F. and {Daruich}, Felipe and {Daubard}, Guillaume and {Daues}, Greg and {Dawson}, William and {Delgado}, Francisco and {Dellapenna}, Alfred and {de Peyster}, Robert and {de Val-Borro}, Miguel and {Digel}, Seth W. and {Doherty}, Peter and {Dubois}, Richard and {Dubois-Felsmann}, Gregory P. and {Durech}, Josef and {Economou}, Frossie and {Eifler}, Tim and {Eracleous}, Michael and {Emmons}, Benjamin L. and {Fausti Neto}, Angelo and {Ferguson}, Henry and {Figueroa}, Enrique and {Fisher-Levine}, Merlin and {Focke}, Warren and {Foss}, Michael D. and {Frank}, James and {Freemon}, Michael D. and {Gangler}, Emmanuel and {Gawiser}, Eric and {Geary}, John C. and {Gee}, Perry and {Geha}, Marla and {Gessner}, Charles J.~B. and {Gibson}, Robert R. and {Gilmore}, D. Kirk and {Glanzman}, Thomas and {Glick}, William and {Goldina}, Tatiana and {Goldstein}, Daniel A. and {Goodenow}, Iain and {Graham}, Melissa L. and {Gressler}, William J. and {Gris}, Philippe and {Guy}, Leanne P. and {Guyonnet}, Augustin and {Haller}, Gunther and {Harris}, Ron and {Hascall}, Patrick A. and {Haupt}, Justine and {Hernandez}, Fabio and {Herrmann}, Sven and {Hileman}, Edward and {Hoblitt}, Joshua and {Hodgson}, John A. and {Hogan}, Craig and {Howard}, James D. and {Huang}, Dajun and {Huffer}, Michael E. and {Ingraham}, Patrick and {Innes}, Walter R. and {Jacoby}, Suzanne H. and {Jain}, Bhuvnesh and {Jammes}, Fabrice and {Jee}, M. James and {Jenness}, Tim and {Jernigan}, Garrett and {Jevremovi{\'c}}, Darko and {Johns}, Kenneth and {Johnson}, Anthony S. and {Johnson}, Margaret W.~G. and {Jones}, R. Lynne and {Juramy-Gilles}, Claire and {Juri{\'c}}, Mario and {Kalirai}, Jason S. and {Kallivayalil}, Nitya J. and {Kalmbach}, Bryce and {Kantor}, Jeffrey P. and {Karst}, Pierre and {Kasliwal}, Mansi M. and {Kelly}, Heather and {Kessler}, Richard and {Kinnison}, Veronica and {Kirkby}, David and {Knox}, Lloyd and {Kotov}, Ivan V. and {Krabbendam}, Victor L. and {Krughoff}, K. Simon and {Kub{\'a}nek}, Petr and {Kuczewski}, John and {Kulkarni}, Shri and {Ku}, John and {Kurita}, Nadine R. and {Lage}, Craig S. and {Lambert}, Ron and {Lange}, Travis and {Langton}, J. Brian and {Le Guillou}, Laurent and {Levine}, Deborah and {Liang}, Ming and {Lim}, Kian-Tat and {Lintott}, Chris J. and {Long}, Kevin E. and {Lopez}, Margaux and {Lotz}, Paul J. and {Lupton}, Robert H. and {Lust}, Nate B. and {MacArthur}, Lauren A. and {Mahabal}, Ashish and {Mandelbaum}, Rachel and {Markiewicz}, Thomas W. and {Marsh}, Darren S. and {Marshall}, Philip J. and {Marshall}, Stuart and {May}, Morgan and {McKercher}, Robert and {McQueen}, Michelle and {Meyers}, Joshua and {Migliore}, Myriam and {Miller}, Michelle and {Mills}, David J.},
        title = "{LSST: From Science Drivers to Reference Design and Anticipated Data Products}",
      journal = {\apj},
         year = 2019,
        month = mar,
       volume = {873},
       number = {2},
          eid = {111},
        pages = {111},
          doi = {10.3847/1538-4357/ab042c},
archivePrefix = {arXiv},
       eprint = {0805.2366},
 primaryClass = {astro-ph},
       adsurl = {https://ui.adsabs.harvard.edu/abs/2019ApJ...873..111I}
}

@ARTICLE{tiwariRMHD,
       author = {{Tiwari}, Vishal and {Chan}, Chi-Ho and {Bogdanovi{\'c}}, Tamara and {Jiang}, Yan-Fei and {Davis}, Shane W. and {Ferrel}, Simon},
        title = "{Radiation Magnetohydrodynamic Simulation of Sub-Eddington Circumbinary Disk around an Equal-mass Massive Black Hole Binary}",
      journal = {\apj},
         year = 2025,
        month = jun,
       volume = {986},
       number = {2},
          eid = {158},
        pages = {158},
          doi = {10.3847/1538-4357/add408},
archivePrefix = {arXiv},
        note  = {Paper~I},
       eprint = {2502.18584},
 primaryClass = {astro-ph.HE},
       adsurl = {https://ui.adsabs.harvard.edu/abs/2025ApJ...986..158T}
}

@ARTICLE{lubow1991a,
       author = {{Lubow}, Stephen H.},
        title = "{A Model for Tidally Driven Eccentric Instabilities in Fluid Disks}",
      journal = {\apj},
         year = 1991,
        month = nov,
       volume = {381},
        pages = {259},
          doi = {10.1086/170647},
       adsurl = {https://ui.adsabs.harvard.edu/abs/1991ApJ...381..259L}
}

@ARTICLE{Lubow1991b,
       author = {{Lubow}, Stephen H.},
        title = "{Simulations of Tidally Driven Eccentric Instabilities with Application to Superhumps}",
      journal = {\apj},
         year = 1991,
        month = nov,
       volume = {381},
        pages = {268},
          doi = {10.1086/170648},
       adsurl = {https://ui.adsabs.harvard.edu/abs/1991ApJ...381..268L}
}

@ARTICLE{Graham_2015,
       author = {{Graham}, Matthew J. and {Djorgovski}, S.~G. and {Stern}, Daniel and {Drake}, Andrew J. and {Mahabal}, Ashish A. and {Donalek}, Ciro and {Glikman}, Eilat and {Larson}, Steve and {Christensen}, Eric},
        title = "{A systematic search for close supermassive black hole binaries in the Catalina Real-time Transient Survey}",
      journal = {\mnras},
         year = 2015,
        month = oct,
       volume = {453},
       number = {2},
        pages = {1562-1576},
          doi = {10.1093/mnras/stv1726},
archivePrefix = {arXiv},
       eprint = {1507.07603},
 primaryClass = {astro-ph.GA},
       adsurl = {https://ui.adsabs.harvard.edu/abs/2015MNRAS.453.1562G}
}

@article{Most_2025PhysRevD.111.L081304,
  title = {Decoupling of a supermassive black hole binary from its magnetically arrested circumbinary accretion disk},
  author = {Most, Elias R. and Wang, Hai-Yang},
  journal = {Phys. Rev. D},
  volume = {111},
  issue = {8},
  pages = {L081304},
  numpages = {10},
  year = {2025},
  month = {Apr},
  publisher = {American Physical Society},
  doi = {10.1103/PhysRevD.111.L081304},
  url = {https://link.aps.org/doi/10.1103/PhysRevD.111.L081304}
}
\bibliographystyle{aasjournal}

\end{CJK*}
\end{document}